\documentclass[lettersize,journal]{IEEEtran}

\usepackage{cite}
\usepackage{amsmath,amssymb,amsfonts}
\usepackage{graphicx}
\usepackage{textcomp}
\usepackage{stfloats}
\usepackage{array}
\usepackage{url}
\usepackage{verbatim}
\usepackage{balance}
\def\BibTeX{{\rm B\kern-.05em{\sc i\kern-.025em b}\kern-.08em
		T\kern-.1667em\lower.7ex\hbox{E}\kern-.125emX}}

\usepackage{subfigure}
\usepackage{epstopdf}

\usepackage{algorithmic}
\usepackage[ruled,vlined]{algorithm2e}

\begin{document}

\include{header}

\title{Integrated Sensing, Communication, and Powering over Multi-antenna OFDM Systems}

\author{Yilong Chen, Chao Hu, Zixiang Ren, Han Hu, Jie Xu, Lexi Xu, Lei Liu, and Shuguang Cui
	\thanks{Y. Chen, C. Hu, Z. Ren, J. Xu, and S. Cui are with the School of Science and Engineering (SSE), the Shenzhen Future Network of Intelligence Institute (FNii-Shenzhen), and the Guangdong Provincial Key Laboratory of Future Networks of Intelligence, The Chinese University of Hong Kong, Shenzhen, Guangdong 518172, China (e-mail: yilongchen@link.cuhk.edu.cn; chaohu@link.cuhk.edu.cn; rzx66@mail.ustc.edu.cn; xujie@cuhk.edu.cn; shuguangcui@cuhk.edu.cn).}
	\thanks{Z. Ren is also with the Key Laboratory of Wireless-Optical Communications, Chinese Academy of Sciences, School of Information Science and Technology, University of Science and Technology of China, Hefei 230027, China.}
	\thanks{H. Hu is with the School of Information and Electronics, Beijing Institute of Technology, Beijing 100081, China (e-mail: hhu@bit.edu.cn).}
	\thanks{L. Xu is with the China Unicom Research Institute, Beijing, China (e-mail: davidlexi@hotmail.com).}
	\thanks{L. Liu is with Anhui Jiaoxin Technology Co., Ltd, Hefei 230041, China (e-mail: liul313@foxmail.com).}
	\thanks{J. Xu is the corresponding author.}
}
\maketitle

\begin{abstract}
	This paper considers a multi-functional orthogonal frequency division multiplexing (OFDM) system with integrated sensing, communication, and powering (ISCAP), in which a multi-antenna base station (BS) transmits OFDM signals to simultaneously deliver information to multiple information receivers (IRs), provide energy supply to multiple energy receivers (ERs), and sense potential targets based on the echo signals.
	To facilitate ISCAP, the BS employs the joint transmit beamforming design by sending dedicated sensing/energy beams jointly with information beams. Furthermore, we consider the beam scanning for sensing, in which the joint beams scan in different directions over time to sense potential targets. In order to ensure the sensing beam scanning performance and meet the communication and powering requirements, it is essential to properly schedule IRs and ERs and design the resource allocation over time, frequency, and space.
	More specifically, we optimize the joint transmit beamforming over multiple OFDM symbols and subcarriers, with the objective of minimizing the average beampattern matching error of beam scanning for sensing, subject to the constraints on the average communication rates at IRs and the average harvested power at ERs.
	We find converged high-quality solutions to the formulated problem by proposing efficient iterative algorithms based on advanced optimization techniques. 
	We also develop various heuristic designs based on the principles of zero-forcing (ZF) beamforming, round-robin user scheduling, and time switching, respectively.
	Numerical results show that our proposed algorithms adaptively generate information and sensing/energy beams at each time-frequency slot to match the scheduled IRs/ERs with the desired scanning beam, significantly outperforming the heuristic designs.
\end{abstract}


\section{Introduction}

Sixth-generation (6G) wireless networks are envisioned to support various intelligent applications such as smart city, smart home, and smart manufacturing, by providing enhanced communication data rate and high positioning accuracy for massive low-power internet-of-things (IoT) devices in a sustainable manner \cite{tong20226g}. Towards this end, 6G networks are expected to integrate advanced radar sensing and wireless power transfer (WPT) functions to enable integrated sensing and communications (ISAC) \cite{cui2021integrating, liu2022Integrated} and wireless information and power transfer (WIPT) \cite{zeng2017communications, clerckx2019fundamentals}, respectively. With the advancements of ISAC and WIPT, integrated sensing, communication, and powering (ISCAP) has emerged as a promising technology to further unify the three functions in 6G networks, in which base stations (BSs) and wireless devices can exploit radio signals to support sensing, communications, and WPT simultaneously, thus significantly enhancing the utilization efficiency of scarce spectrum, energy, and hardware resources \cite{chen2024integrated}.

While ISAC and WIPT have been extensively studied independently (see, e.g., \cite{hua2024mimo, ren2024fundamental, xu2014multiuser} and the references therein), the research of ISCAP is still in its infancy. \cite{chen2024isac} is the initial work investigating ISCAP in a multiple-input multiple-output (MIMO) setup, in which a multi-functional BS utilizes unified signals for simultaneously communicating with an information receiver (IR), wirelessly powering an energy receiver (ER), and sensing a target. The authors in \cite{chen2024isac} derived the optimal transmit covariance solution in a semi-closed form to optimally balance the performance tradeoff among the three functions of sensing, communication, and WPT. The solution exhibits an interesting structure, which unifies the rate-maximization-oriented eigenmode transmission with water-filling power allocation for communication, the energy-maximization-oriented strongest eigenmode transmission for WPT, and the Cram{\'e}r–Rao bound (CRB)-minimization-oriented design for target sensing (e.g., isotropic transmission for the case with extended target estimation).
Next, the follow-up work \cite{zhou2023integrating} extended the MIMO ISACP design to the multi-antenna setup with multiple ERs and IRs. Furthermore, \cite{hao2024energy} studied the energy-efficient multi-antenna ISCAP with hybrid analog-and-digital beamforming at the multi-functional BS by considering a comprehensive energy model for the BS, in which dedicated sensing/energy beams are jointly transmitted in addition to information beams to provide full degrees of freedom (DoFs) for sensing and WPT. In addition, \cite{zhang2024integrated} exploited the massive MIMO for ISCAP, \cite{ren2024secure} utilized extremely large-scale antenna arrays for secure ISCAP by employing the beam-focusing effect in the near field, and \cite{yang2024joint} studied the intelligent reflecting surface (IRS)-assisted ISCAP. Besides the simultaneous multi-functional transmissions, there have been some other prior works investigating the interplay among the three functions via, e.g., integrated sensing and powering \cite{li2023intelligent, mayer2023joint}, wireless powered ISAC \cite{li2022wirelessly}, and sensing-assisted WPT \cite{zhang2023training} or WIPT \cite{xu2024sensing}. However, these prior works focused on the ISCAP over narrowband channels, which may not work well for practical wireless systems with wideband transmission.

Different from the above works, this paper studies the ISCAP over a multi-antenna orthogonal frequency division multiplexing (OFDM) system, in which a multi-antenna BS performs joint information and sensing/energy beamforming to simultaneously deliver information to multiple IRs, provide energy supply to multiple ERs, and estimate potential sensing targets based on the echo signals.
In particular, we consider a practical beam scanning scheme for radar sensing \cite{zhang2019multibeam, pucci2022system}, in which the signal beams need to scan different directions over time to sense potential targets. Accordingly, we utilize the matching error between the transmit beampattern and the desired beampattern over time as the sensing performance metric \cite{fuhrmann2008transmit, liu2018mumimo}. In this case, it is important to perform the user scheduling and resource allocation over time, frequency, and space domains for balancing the performance tradeoff among beam scanning for sensing, WPT for ERs, and communications with IRs. 
This task, however, is challenging due to the following technical issues. First, 
%due to the involvement of three domains, the dimension of resource allocation variables becomes high, thus making the optimization design rather difficult. Next, 
the inter-user interference plays different roles for the three functions in the multiuser ISCAP system. In particular, the interference is detrimental to communication but can be beneficial for powering and sensing \cite{chen2024integrated}.
%, thus making it a crucial but complicated problem to handle. In addition, the beamforming design principles are different for the three functions. 
To be specific, for powering and sensing, both the information and dedicated sensing/energy signals are harvestable by ERs, and exploitable for beampattern matching,
%can be jointly exploited to match the beampatterns with the desired time-varying sensing directions, 
respectively; while for communications, only the information signals are desired by IRs, and the signals for other IRs and sensing/energy signals become harmful interference that should be properly mitigated. This thus makes it a crucial but complicated problem to design the joint beamforming and handle the interference. Next, different IRs and ERs can be scheduled at different time-frequency slots to match the beam scanning directions over time and facilitate the co-channel interference management. Due to the scheduling and resource allocation over the time, frequency, and space domains, the dimension of decision variables becomes significantly high, thus making the optimization design rather difficult. Therefore, we are motivated to address these issues in this work.

The main results of this paper are summarized as follows. 

\begin{itemize}
	\item 
	To provide full DoFs for ISCAP, we employ the joint information and dedicated sensing/energy beamforming to support beam scanning for sensing, communication with IRs, and WPT for ERs. 
	To characterize the fundamental performance tradeoff among the three functions, we optimize the joint beamforming over OFDM symbols and subcarriers, with the objective of minimizing the beampattern matching error for sensing beam scanning, subject to the communication rate requirements at IRs, the harvested power requirements at ERs, and the transmit power constraint at the BS. The formulated beampattern matching error minimization problem is highly non-convex, and thus is difficult to solve. 
	\item First, we propose converged solutions with high quality to tackle this problem. Towards this end, we first utilize the semi-definite relaxation (SDR) technique to relax the non-convex problem, and then adopt two techniques, namely the successive convex approximation (SCA) and fractional programming (FP), to find converged solutions to the relaxed problems.
%	solve the relaxed problem in an iterative manner by handling the rate constraints at IRs. In each iteration of the SCA and FP-based algorithms, we use the convex optimization technique to solve the reformulated problems with the same quadratic objective but different rate constraints for the SCA and FP-based algorithms, which lead to the same converged solution with different complexities.
%	Furthermore, the complexity and convergency of our proposed solutions are analyzed.
	\item Next, we present three alternative heuristic designs to solve the considered beampattern matching error minimization problem based on the principles of zero-forcing (ZF) beamforming, round-robin user scheduling, and time switching, respectively. In general, these heuristic solutions can be implemented with lower complexity and/or serve as performance benchmarks. 
	\item Finally, numerical results show that our proposed joint beamforming designs based on SDR and SCA/FP significantly outperform other heuristic benchmarks by adaptively scheduling IRs and ERs in the time-frequency-space domains to match the scanning beam for sensing. This thus verifies the superiority of our proposed joint beamforming designs. Furthermore, we verify the effectiveness of our proposed designs in balancing the performance tradeoff among the three functions of sensing, communication, and WPT. In addition, the sensing performance is evaluated by considering a practical target estimation task.
%	beampattern matching error minimization design by showing the practical target estimation performance.
%	In numerical results, the outperformance over these benchmarks under different system setups verifies the superiority of our proposed joint beamforming design, which adaptively schedules IRs and ERs in the time-frequency-space domain to match the scanning beam for sensing.
%	Furthermore, we verify the effectiveness of our proposed joint beamforming design in target estimation based on corresponding receiver algorithms.
\end{itemize}

To our best knowledge, there is only prior work \cite{zhang2023multi} that investigated the ISCAP over OFDM systems by considering the single-antenna scenario with one single IR and ER (instead of multi-antenna setup with multiple IRs and ERs in this paper). In \cite{zhang2023multi}, the non-zero mean asymmetric Gaussian distributed signal input was considered for performing the three functions at the same time. By contrast, this paper considers a more complicated multiuser multi-antenna setup by employing joint information and sensing/energy beamforming designs, via considering a practical beam scanning scheme for sensing.

The remainder of this paper is organized as follows. Section II presents the multi-antenna OFDM ISCAP system model and formulates the joint beamforming design problem. Section III proposes efficient solutions to the formulated problem by using SDR and SCA/FP. Section IV develops heuristic designs based on the principles of ZF beamforming, round-robin user scheduling, and time switching, respectively. Section V provides numerical results to validate the performance of our proposed designs. Finally, Section VI concludes this paper.

\textit{Notations:} Boldface letters are used for vectors (lower-case) and matrices (upper-case). For a square matrix \(\mathbf{A}\), \(\mathrm{tr}(\mathbf{A})\) and \(\mathbf{A}^{-1}\) denote its trace and inverse, respectively, while \(\mathbf{A} \succeq \mathbf{0}\) means that \(\mathbf{A}\) is positive semidefinite. For an arbitrary-size matrix \(\mathbf{A}\), \(\mathrm{rank}(\mathbf{A})\), \(\mathbf{A}^*\), \(\mathbf{A}^T\), and \(\mathbf{A}^H\) denote its rank, conjugate, transpose, and conjugate transpose, respectively, and \([\mathbf{A}]_{:,n}\) denotes its \(n\)-th column. For a vector \(\mathbf{a}\), \(\|\mathbf{a}\|\) denotes its Euclidean norm. For a complex number \(a\), \(|a|\) denotes its magnitude. 
\(\mathrm{diag}(a_1, \dots, a_N)\) denotes a diagonal matrix whose diagonal entries are \(a_1, \dots, a_N\).
\(\mathrm{mod}(a,b)\) denotes the modulo operator, where integers \(a\) and \(b\) are the dividend and divisor, respectively.
\(\mathbf{I}_N\) denotes the identity matrix with dimension \(N \times N\). \(\mathbb{C}^{M \times N}\) denotes the space of \(M \times N\) complex matrices. \(\mathbb{E}[\cdot]\) denotes the statistic expectation. 
\(j = \sqrt{-1}\) denotes the imaginary unit. 

\section{System Model}

\begin{figure}[tb]
	\centering {\includegraphics[width=0.5\textwidth]{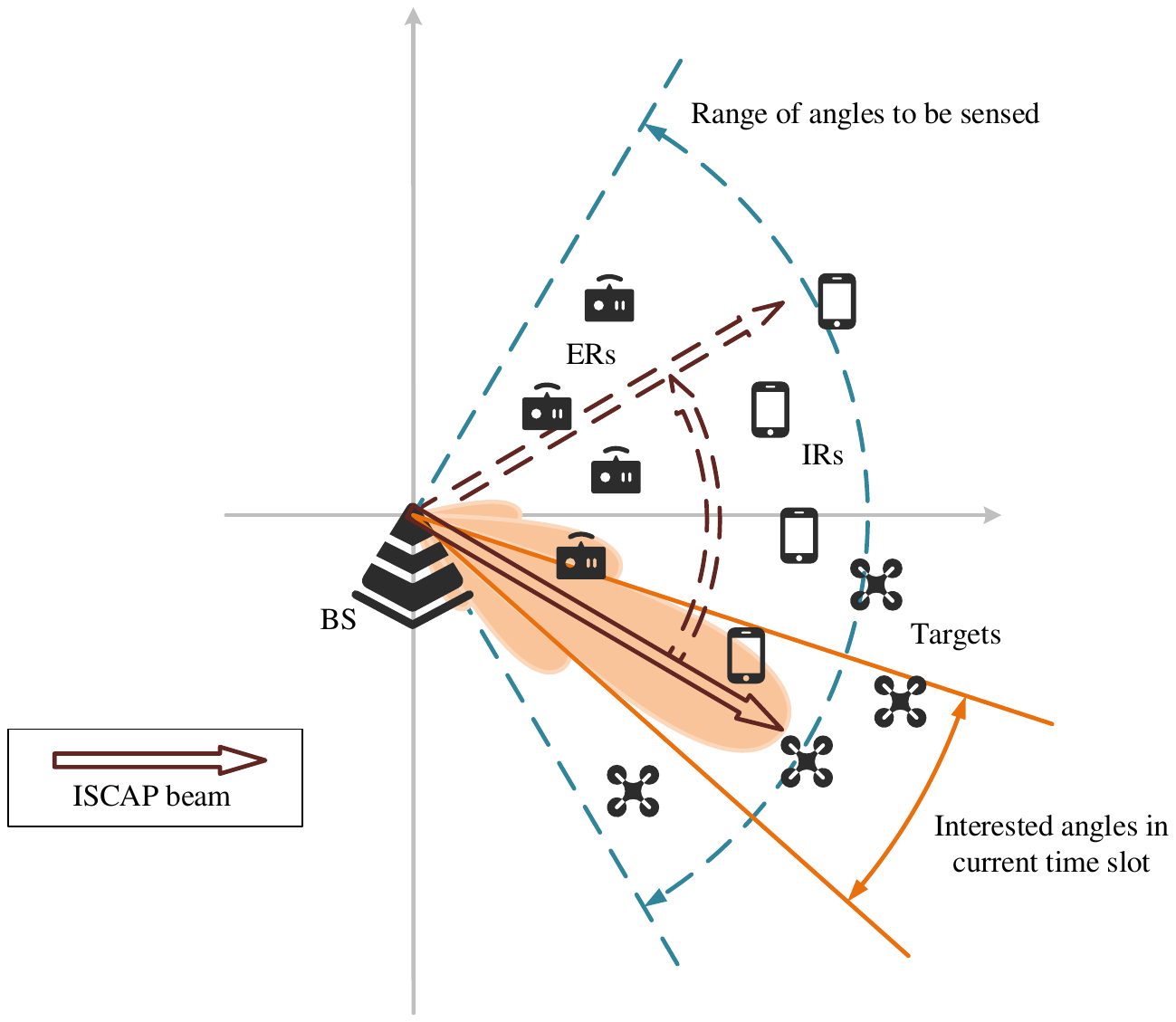}} 
	\caption{The proposed ISCAP system with beam scanning.}
	\label{Fig_system}
\end{figure}
We consider a downlink multi-antenna OFDM ISCAP system, as illustrated in Fig. \ref{Fig_system}, where a multi-antenna BS transmits wireless signals to simultaneously deliver information to \(K_\text{IR} \ge 1\) IRs, provide power supply to \(K_\text{ER} \ge 1\) ERs, and sense potential objects over a targeted area. Let \(\mathcal{K}_\text{IR} \triangleq \{1, \dots, K_\text{IR}\}\) and \(\mathcal{K}_\text{ER} \triangleq \{1, \dots, K_\text{ER}\}\) denote the sets of IRs and ERs, respectively. The BS is assumed to be equipped with a uniform linear array (ULA) of \(N_t > 1\) transmit antennas and \(N_r \ge N_t\) receive antennas for sensing,\footnote{For radar estimation tasks, the number of receive antennas should be generally greater than that of transmit antennas to prevent any potential information loss of the sensed targets \cite{chen2024isac, hua2024mimo}.}
%	Additionally, the BS has to operate in full-duplex mode to transmit signals and receive echo signals at the same time.
while IRs and ERs are each equipped with a single receive antenna. To reveal the fundamental performance upper bound limits, it is assumed that the BS has perfect channel state information (CSI), as commonly assumed in the literature, e.g., \cite{chen2024isac, hua2024mimo, ren2024fundamental, xu2014multiuser}.

We focus on the ISCAP transmission over a set \(\mathcal{L} \triangleq \{0, \dots, L-1\}\) of \(L\) OFDM symbols, where each OFDM symbol consists of a set \(\mathcal{N} \triangleq \{0, \dots, N-1\}\) of \(N\) subcarriers. We denote the duration of each OFDM symbol as \(T_\text{sym}\) and the bandwidth of all \(N\) subcarriers by \(B = N \Delta_f\), where \(\Delta_f\) is the subcarrier spacing. The carrier frequency is denoted as \(f_c\).
Furthermore, we divide the angle range \([-\frac{\pi}{2},\frac{\pi}{2}]\) into a set of quantized angular grids \(\Theta \triangleq \big\{\theta_m | m \in \mathcal{M} \triangleq \{1, \dots, M\}\big\}\).

As shown in Fig. 1, we consider the beam scanning for target sensing, where a time-varying directional beam is employed by the BS, steering towards interested angles over time. The beam scanning scheme allows the BS to sense over a wide area, and the accurately steered beam at each time helps to achieve a large beamforming gain in the direction of interest while reducing the effect of clutter from other directions \cite{zhang2019multibeam, pucci2022system}.
Since the sensing process needs to last for a specified duration, we divide the transmitted symbols in \(\mathcal{L}\) into a set \(\mathcal{Q} = \{1,\dots,Q\}\) of \(Q\) time slots with equal length, denoted by \(\mathcal{L}_1, \dots, \mathcal{L}_Q\).
During beam scanning, in each time slot \(q \in \mathcal{Q}\) with symbols in \(\mathcal{L}_q\), the BS is interested in a specific set of angles denoted by \(\Theta_q \subseteq \Theta\).
We assume that the angles in \(\Theta_q\) monotonically increase as \(q\) progresses from \(1\) to \(Q\), and \(\cup_{q\in\mathcal{Q}} \Theta_q = \Theta\) covers the entire range of angles to be sensed.

To facilitate the ISCAP, the BS sends information and dedicated sensing/energy signals \cite{liu2018mumimo, liuxiang2020joint}.
Let \(s_{n,l,k}\) denote the information signal for IR \(k\in\mathcal{K}_\text{IR}\) at OFDM symbol \(l\in\mathcal{L}\) and subcarrier \(n\in\mathcal{N}\), and \(\mathbf{w}_{n,l,k} \in \mathbb{C}^{N_t \times 1}\) denote the corresponding transmit beamforming vector. Here, \(s_{n,l,k}\) is assumed to be a circularly symmetric complex Gaussian (CSCG) random variable with zero mean and unit variance. For notational convenience, we define \(\mathbf{W}_{n,l,k} \triangleq \mathbf{w}_{n,l,k} \mathbf{w}_{n,l,k}^H\) with \(\mathbf{W}_{n,l,k} \succeq \mathbf{0}\) and \(\mathrm{rank}(\mathbf{W}_{n,l,k}) \le 1\), \(\forall n\in\mathcal{N}, l\in\mathcal{L}, k \in \mathcal{K}_\text{IR}\). Accordingly, the optimization of beamforming vector \(\mathbf{w}_{n,l,k}\) is equivalent to optimizing \(\mathbf{W}_{n,l,k}\). 
Furthermore, let \(\mathbf{s}_{n,l,0} \in \mathbb{C}^{N_t \times 1}\) denote the dedicated sensing/energy signal at OFDM symbol \(l\in\mathcal{L}\) and subcarrier \(n\in\mathcal{N}\), which is assumed to be a pseudorandom sequence with zero mean and covariance matrix \(\mathbb{E}[\mathbf{s}_{n,l,0} \mathbf{s}_{n,l,0}^H] = \mathbf{W}_{n,l,0} \succeq \mathbf{0}\).
The transmitted signal at each OFDM symbol \(l\in\mathcal{L}\) and subcarrier \(n\in\mathcal{N}\) is accordingly expressed as
\begin{equation}
	\mathbf{x}_{n,l} = \sum_{k\in\mathcal{K}_\text{IR}} \mathbf{w}_{n,l,k} s_{n,l,k} + \mathbf{s}_{n,l,0}.
\end{equation}
Consequently, the covariance matrix of \(\mathbf{x}_{n,l}\) is given by
\begin{equation}
	\mathbf{R}_{n,l} = \mathbb{E}[\mathbf{x}_{n,l} \mathbf{x}_{n,l}^H] = \sum_{k\in\{0\}\cup\mathcal{K}_\text{IR}} \mathbf{W}_{n,l,k}.
\end{equation}
Let \(P_0\) denote the maximum transmit power at the BS. We thus have the transmit power constraint as
\begin{equation} \label{P0}
	\sum_{n\in\mathcal{N}} \sum_{k\in\{0\}\cup\mathcal{K}_\text{IR}} \mathrm{tr}(\mathbf{W}_{n,l,k}) = P_0, \forall l\in\mathcal{L}.
\end{equation}
In \eqref{P0}, we consider the transmit power to be strictly equal to the maximum value \(P_0\) to ensure the sensing performance, similarly as in MIMO radar sensing literature \cite{liu2018mumimo, liuxiang2020joint}. 

\subsection{Sensing Model}

In this subsection, we focus on the target sensing with beam scanning. We use the beampattern matching error \cite{fuhrmann2008transmit, liu2018mumimo} as the sensing performance metric,
%\footnote{In Section V, we will evaluate the performance of the transmit beamforming design by considering specific target estimation methods.} 
which is defined as the squared error between the transmit beampattern gain and the desired beampattern gain.

We first define the transmit beampattern gain as follows, which generally depicts the transmit signal power distribution at each OFDM symbol \(l\in\mathcal{L}\) with respect to (w.r.t.) each angle \(\theta \in \Theta\).\footnote{Here, the beampattern gain is summed over the \(N\) subcarriers. This consideration is consistent with the target angle estimation process, in which the received signals over the \(N\) subcarriers are jointly exploited, as shown in Appendix B.}
\begin{equation}
	\mathcal{G}_l\big(\theta, \{\mathbf{W}_{n,l,k}\}\big) = \sum_{n\in\mathcal{N}} \sum_{k\in\{0\}\cup\mathcal{K}_\text{IR}} \mathbf{v}^T(\theta) \mathbf{W}_{n,l,k} \mathbf{v}^*(\theta),
\vspace{-0.1cm}\end{equation}
where \(\mathbf{v}(\theta) = [1, e^{j2\pi\frac{d}{\lambda}\sin\theta}, \dots, e^{j2\pi\frac{d}{\lambda}(N_t-1)\sin\theta}]^T\) denotes the steering vector at angle \(\theta\), with \(\lambda\) and \(d\) denoting the carrier wavelength and the spacing between two adjacent antennas, respectively.

In addition, for each OFDM symbol \(l\in\mathcal{L}\) and angle \(\theta \in \Theta\), we denote the desired beampattern gain as \(\mathcal P_l(\theta)\). To mitigate the interference of scatters in uninterested directions, \(\mathcal{P}_l(\theta)\) is defined as a rectangular function w.r.t. the interested angles, i.e.,
\begin{equation}
	\mathcal{P}_l(\theta) =
	\begin{cases}
		1, & \theta \in \Theta_q, l\in\mathcal{L}_q \\
		0, & \theta \notin \Theta_q, l\in\mathcal{L}_q 
	\end{cases}.
\end{equation}

Accordingly, over the \(L\) OFDM symbols, the beampattern matching error between the transmit beampattern gain and the desired beampattern gain across the angular grid \(\Theta\) is expressed as
\begin{equation} \label{err}
	\mathcal{E}\big(\{\mathbf{W}_{n,l,k}\}, \zeta\big) = \sum_{l\in\mathcal{L}} \sum_{m \in \mathcal{M}} \Big|\mathcal{G}_l\big(\theta_m, \{\mathbf{W}_{n,l,k}\}\big) - \zeta \mathcal{P}_l(\theta_m)\Big|^2,
\end{equation}
where \(\zeta\) is a scaling coefficient to be designed. 

\subsection{Communication Model}

Next, we consider the communication with multiple IRs. At each IR \(k\in\mathcal{K}_\text{IR}\), the received signal at OFDM symbol \(l\in\mathcal{L}\) and subcarrier \(n\in\mathcal{N}\) is expressed as
\begin{equation}
	\begin{aligned}
		y_{n,l,k}^\text{IR} =& \mathbf{h}_{n,k}^H \mathbf{x}_{n,l} + z_{n,l,k} \\
		=& \underbrace{\mathbf{h}_{n,k}^H \mathbf{w}_{n,l,k} s_{n,l,k}}_{\text{information signal}} + \underbrace{\mathbf{h}_{n,k}^H \sum_{i\in\mathcal{K}_\text{IR}\setminus\{k\}} \mathbf{w}_{n,l,i} s_{n,l,i}}_{\text{inter-user interference}} \\
		&+ \underbrace{\mathbf{h}_{n,k}^H \mathbf{s}_{n,l,0}}_{\text{co-channel interference}} + \underbrace{z_{n,l,k}}_{\text{noise}},
	\end{aligned}
\end{equation}
where \(\mathbf{h}_{n,k}^H \in \mathbb{C}^{1\times N_t}\) denotes the channel vector from the BS to IR \(k\) at subcarrier \(n\), and \(z_{n,l,k}\) denotes the noise at the receiver of IR \(k\), which is a CSCG random variable with zero mean and variance \(\sigma_c^2\).
The corresponding signal-to-interference-plus-noise ratio (SINR) at IR \(k\) over OFDM symbol \(l\) and subcarrier \(n\) is given by
\begin{equation}
	\gamma_{n,l,k}\big(\{\mathbf{W}_{n,l,k}\}\big) = \frac{\mathbf{h}_{n,k}^H \mathbf{W}_{n,l,k} \mathbf{h}_{n,k}}{\sum_{i\in\{0\}\cup\mathcal{K}_\text{IR}\setminus\{k\}} \mathbf{h}_{n,k}^H \mathbf{W}_{n,l,i} \mathbf{h}_{n,k} + \sigma_c^2}.
\end{equation}
With Gaussian signaling, the achievable average data rate (in bps/Hz) over the \(L\) symbols and \(N\) subcarriers at each IR \(k\in\mathcal{K}_\text{IR}\) is given by \cite{telatar1999capacity}
\begin{equation} \label{rate}
	\begin{aligned}
		R_k\big(\{\mathbf{W}_{n,l,k}\}\big) &= \frac{1}{LN} \sum_{l\in\mathcal{L}} \sum_{n\in\mathcal{N}} \log_2\Big(1+ \gamma_{n,l,k}\big(\{\mathbf{W}_{n,l,k}\}\big)\Big).
	\end{aligned}
\end{equation}

\subsection{Powering Model}

Then, we consider the WPT towards ERs. Each ER adopts the rectifiers to convert the received radio frequency (RF) signals into direct current (DC) signals for energy harvesting. In general, the harvested DC power is monotonically non-decreasing w.r.t. the received RF power. As a result, we use the received RF power (energy over a unit time period, in Watt) at each ER as the performance metric for powering.\footnote{When the RF-to-DC energy conversion process of the rectifier is linear, maximizing the harvested RF power is equivalent to maximizing the harvested DC power. In practice, this process may be non-linear and the harvested DC power may be modeled by considering the non-linear diode current-voltage characteristics and the DC power saturation at large RF power due to the diode breakdown (see, e.g., \cite{clerckx2019fundamentals}). However, such non-linear models are out of the technical scope of this paper.}
At each ER \(i\in\mathcal{K}_\text{ER}\), the average received RF power over the \(L\) OFDM symbols and \(N\) subcarriers is given by \cite{clerckx2019fundamentals}
\begin{equation} \label{RFp}
	P_{i}^\text{ER}\big(\{\mathbf{W}_{n,l,k}\}\big) = \frac{1}{L} \sum_{l\in\mathcal{L}} \sum_{n\in\mathcal{N}} \sum_{k\in\{0\}\cup\mathcal{K}_\text{IR}} \mathbf{g}_{n,i}^H \mathbf{W}_{n,l,k} \mathbf{g}_{n,i},
\end{equation}
where \(\mathbf{g}_{n,i}^H \in \mathbb{C}^{1\times N_t}\) denotes the channel vector from the BS to ER \(i\) over subcarrier \(n\in\mathcal{N}\).

\subsection{Problem Formulation}

Our objective is to optimize the joint transmit beamforming \(\{\mathbf{W}_{n,l,k}\}\) across OFDM symbols and subcarriers to minimize the beampattern matching error \(\mathcal{E}\big(\{\mathbf{W}_{n,l,k}\}, \zeta\big)\) in \eqref{err} for sensing beam scanning, while ensuring that the average communication rate \(R_k\big(\{\mathbf{W}_{n,l,k}\}\big)\) in \eqref{rate} at each IR \(k \in \mathcal{K}_\text{IR}\) and the average harvested power \(P_{i}^\text{RF}\big(\{\mathbf{W}_{n,l,k}\}\big)\) in \eqref{RFp} at each ER \(i \in \mathcal{K}_\text{ER}\) meet the corresponding requirements \(\Gamma_\text{IR}\) and \(\Gamma_\text{ER}\), respectively, subject to the transmit power constraint at the BS. Mathematically, the problem is formulated as
	\begin{subequations}
		\begin{align} 
			\text{(P1):} \min_{\{\mathbf{W}_{n,l,k}\}, \zeta} &\ \mathcal{E}\big(\{\mathbf{W}_{n,l,k}\}, \zeta\big) \label{1s} \\
			\mathrm{s.t.} \quad &\ R_k\big(\{\mathbf{W}_{n,l,k}\}\big) \ge \Gamma_\text{IR}, \forall k \in \mathcal{K}_\text{IR}, \label{1c} \\
			&\ P_{i}^\text{ER}\big(\{\mathbf{W}_{n,l,k}\}\big) \ge \Gamma_\text{ER}, \forall i \in \mathcal{K}_\text{ER}, \label{1p} \\ 
			&\ \sum_{n\in\mathcal{N}} \sum_{k\in\{0\}\cup\mathcal{K}_\text{IR}} \mathrm{tr}(\mathbf{W}_{n,l,k}) = P_0, \forall l \in \mathcal{L}, \label{1b} \\
			&\ \mathbf{W}_{n,l,k} \succeq \mathbf{0}, \forall n\in\mathcal{N}, l\in\mathcal{L}, k\in\{0\}\cup\mathcal{K}_\text{IR}, \label{1sdp} \\
			&\ \mathrm{rank}(\mathbf{W}_{n,l,k}) \le 1, \forall n\in\mathcal{N}, l\in\mathcal{L}, k\in\mathcal{K}_\text{IR}. \label{1one}
		\end{align}
	\end{subequations}
Intuitively, optimizing the joint beamforming \(\{\mathbf{W}_{n,l,k}\}\) across OFDM symbols and subcarriers in problem (P1) involves the time-frequency-space scheduling of the IRs and ERs. This ensures that the communication/powering overhead at each IR/ER dynamically aligns with the variation of the scanning beam to optimize the sensing performance.
However, problem (P1) is non-convex, making it challenging to be optimally solved in general. The non-convexity arises from the communication rate constraints in \eqref{1c} and the rank-one constraints in \eqref{1one}. In the following, we focus on solving problem (P1).

\section{Proposed Solutions to Problem (P1)}

In this section, we propose efficient algorithms to solve the non-convex problem (P1). In the following, we first relax the rank-one constraints in \eqref{1one} by using the SDR technique, and then present two algorithms based on SCA and FP, respectively, to iteratively refine the solution towards convergence.
%The two designs are widely adopted, and we would like to consider both of them and compare their performance and complexity.

First, by utilizing SDR \cite{luo2010semidefinite} to remove the rank-one constraints in  \eqref{1one}, problem (P1) is relaxed to the following form.
\begin{equation*}
	\begin{aligned} 
		\text{(P2):} \min_{\{\mathbf{W}_{n,l,k}\}, \zeta} &\ \mathcal{E}\big(\{\mathbf{W}_{n,l,k}\}, \zeta\big) \\
		\mathrm{s.t.} \quad &\ \eqref{1c}, \ \eqref{1p}, \ \eqref{1b}, \ \text{and} \ \eqref{1sdp}.
	\end{aligned}
\end{equation*}
Suppose that the obtained solution to problem (P2) is given by \(\{\hat{\mathbf{W}}_{n,l,k}\}\) and \(\hat{\zeta}\), which will be specified later. Given \(\{\hat{\mathbf{W}}_{n,l,k}\}\) and \(\hat{\zeta}\), we need to find a feasible solution to the original problem (P1).
Notice that if the solution \(\{\hat{\mathbf{W}}_{n,l,k}\}\) to (P2) satisfies \(\mathrm{rank}(\hat{\mathbf{W}}_{n,l,k}) \le 1\),  \(\forall n\in\mathcal{N}, l\in\mathcal{L}, k\in\mathcal{K}_\text{IR}\), then it is also feasible and efficient to (P1). 
Fortunately, as shown in the following proposition, even if (P2) yields a high-rank solution, there always exists an equivalent rank-one solution to achieve the same objective value.

\textbf{Proposition 1:} Given solution \(\{\hat{\mathbf{W}}_{n,l,k}\}\) and \(\hat{\zeta}\) to problem (P2), we can always construct an alternative solution \(\{\bar{\mathbf{W}}_{n,l,k}\}\) and \(\bar{\zeta}\) in the following with \(\mathrm{rank}(\bar{\mathbf{W}}_{n,l,k}) \le 1\), \(\forall n\in\mathcal{N}, l\in\mathcal{L}, k\in\mathcal{K}_\text{IR}\), which is feasible for (P1) and achieves the same objective value as that of (P2) achieved by \(\{\hat{\mathbf{W}}_{n,l,k}\}\) and \(\hat{\zeta}\).
\begin{subequations}
	\begin{align}
		&\bar{\zeta} = \hat{\zeta}, \label{trans1}\\
		&\bar{\mathbf{W}}_{n,l,k} = \bar{\mathbf{w}}_{n,l,k} \bar{\mathbf{w}}_{n,l,k}^H, \forall k\in\mathcal{K}_\text{IR}, \ \text{with}\ \nonumber\\ & \qquad  \bar{\mathbf{w}}_{n,l,k} = (\mathbf{h}_{n,k}^H \hat{\mathbf{W}}_{n,l,k} \mathbf{h}_{n,k})^{-\frac{1}{2}} \hat{\mathbf{W}}_{n,l,k} \mathbf{h}_{n,k}, \label{trans2}\\
		&\bar{\mathbf{W}}_{n,l,0} = \sum_{k\in\{0\}\cup\mathcal{K}_\text{IR}} \hat{\mathbf{W}}_{n,l,k} - \sum_{k\in\mathcal{K}_\text{IR}} \bar{\mathbf{W}}_{n,l,k}. \label{trans3}
	\end{align}
\end{subequations}

\textit{Proof:} See Appendix A.
\hfill \(\square\)

According to Proposition 1, problem (P1) can be solved by equivalently solving the relaxed problem (P2).
%based on the obtained solution from problem (P2), we can find a feasible solution for the original problem (P1) that achieves the same objective value. 
In the following, we focus on solving problem (P2), which, however, is still non-convex due to the communication rate constraints in \eqref{1c}. In the following two subsections, we present two algorithms based on the celebrated techniques of SCA and FP, respectively, to solve (P2) in iterative manners. 
%In each iteration of the SCA and FP-based algorithms, we solve convex semi-definite programs with the same quadratic objective but different transformed rate constraints. The convergence and complexity performances of the two algorithms are compared in Section V.

\subsection{SCA-based Algorithm for Solving (P2)}

In this subsection, we employ SCA \cite{sun2017majorization} to tackle problem (P2).
To facilitate the solution, we rewrite (P2) as
\begin{subequations} \label{1.1}
	\begin{align} 
		\min_{\{\mathbf{W}_{n,l,k}\}, \zeta} &\ \mathcal{E}\big(\{\mathbf{W}_{n,l,k}\}, \zeta\big) \\
		\mathrm{s.t.} \quad &\ \sum_{l\in\mathcal{L}} \sum_{n\in\mathcal{N}} \Big(\log_2\big(\sum_{i\in\{0\}\cup\mathcal{K}_\text{IR}} \mathbf{h}_{n,k}^H \mathbf{W}_{n,l,i} \mathbf{h}_{n,k} + \sigma_c^2\big) \nonumber\\
		&\ -\log_2\big(\sum_{i\in\{0\}\cup\mathcal{K}_\text{IR}\setminus\{k\}} \mathbf{h}_{n,k}^H \mathbf{W}_{n,l,i} \mathbf{h}_{n,k} + \sigma_c^2\big)\Big) \nonumber \\ 
		&\qquad \qquad \qquad \qquad \qquad \ \ge LN \Gamma_\text{IR}, \forall k\in\mathcal{K}_\text{IR}, \label{1.1c} \\
		&\ \eqref{1p}, \ \eqref{1b}, \ \text{and} \ \eqref{1sdp},
	\end{align}
\end{subequations}
in which the constraints in \eqref{1.1c} are non-convex. To deal with this issue, we introduce auxiliary variables \(\widetilde{\mathbf{W}}_{n,l,k} = \sum_{i\in\{0\}\cup\mathcal{K}_\text{IR}\setminus\{k\}} \mathbf{W}_{n,l,i}\), \(\forall n\in\mathcal{N}, l\in\mathcal{L}, k\in\mathcal{K}_\text{IR}\), and accordingly re-express \eqref{1.1c} as
\begin{equation} \label{1.1cc}
	\begin{aligned}
		& \sum_{l\in\mathcal{L}} \sum_{n\in\mathcal{N}} \Big(\log_2\big(\mathbf{h}_{n,k}^H (\mathbf{W}_{n,l,k} + \widetilde{\mathbf{W}}_{n,l,k}) \mathbf{h}_{n,k} + \sigma_c^2\big) \\
		& \underbrace{-\log_2\big(\mathbf{h}_{n,k}^H \widetilde{\mathbf{W}}_{n,l,k} \mathbf{h}_{n,k} + \sigma_c^2\big)}_{\triangleq f_{n,l,k}(\widetilde{\mathbf{W}}_{n,l,k})}\Big) \ge LN \Gamma_\text{IR}, \forall k\in\mathcal{K}_\text{IR}.
	\end{aligned}
\end{equation}

It is observed that the non-convexity of \eqref{1.1cc} is due to the terms \(\big\{f_{n,l,k}(\widetilde{\mathbf{W}}_{n,l,k})\big\}\), and thus we use SCA to approximate them into a series of linear forms. Specifically, in each iteration \(j\) of the SCA-based algorithm, let \(\{\mathbf{W}_{n,l,k}^{(j)}\}\) and \(\{\widetilde{\mathbf{W}}_{n,l,k}^{(j)}\}\) denote the current points of \(\{\mathbf{W}_{n,l,k}\}\) and \(\{\widetilde{\mathbf{W}}_{n,l,k}\}\), respectively. According to the first-order Taylor approximation based on local point \(\widetilde{\mathbf{W}}_{n,l,k}^{(j)}\), we approximate \(f_{n,l,k}(\widetilde{\mathbf{W}}_{n,l,k})\) as its linear lower bound in the following, \(\forall n\in\mathcal{N}, l\in\mathcal{L}, k\in\mathcal{K}_\text{IR}\).
\begin{equation} \label{SCA-C0}
	\begin{aligned}
		f_{n,l,k}(\widetilde{\mathbf{W}}_{n,l,k}) \ge &- \log_2\big(\mathbf{h}_{n,k}^H \widetilde{\mathbf{W}}_{n,l,k}^{(j)} \mathbf{h}_{n,k} + \sigma_c^2\big) \\
		&- \frac{\mathbf{h}_{n,k}^H (\widetilde{\mathbf{W}}_{n,l,k} - \widetilde{\mathbf{W}}_{n,l,k}^{(j)}) \mathbf{h}_{n,k}}{(\mathbf{h}_{n,k}^H \widetilde{\mathbf{W}}_{n,l,k}^{(j)} \mathbf{h}_{n,k} + \sigma_c^2) \ln2}.
	\end{aligned}
\end{equation}
Accordingly, we approximate the constraints in \eqref{1.1c} as
\begin{equation} \label{SCA-C}
	\begin{aligned}
		& \sum_{l\in\mathcal{L}} \sum_{n\in\mathcal{N}} \Big(\log_2\big(\sum_{i\in\{0\}\cup\mathcal{K}_\text{IR}} \mathbf{h}_{n,k}^H \mathbf{W}_{n,l,i} \mathbf{h}_{n,k} + \sigma_c^2\big) \\
		&- \log_2\big(\sum_{i\in\{0\}\cup\mathcal{K}_\text{IR}\setminus\{k\}} \mathbf{h}_{n,k}^H \mathbf{W}_{n,l,i}^{(j)} \mathbf{h}_{n,k} + \sigma_c^2\big) \\
		&- \frac{\sum_{i\in\{0\}\cup\mathcal{K}_\text{IR}\setminus\{k\}} \mathbf{h}_{n,k}^H (\mathbf{W}_{n,l,i} - \mathbf{W}_{n,l,i}^{(j)}) \mathbf{h}_{n,k}^H}{(\sum_{i\in\{0\}\cup\mathcal{K}_\text{IR}\setminus\{k\}} \mathbf{h}_{n,k}^H \mathbf{W}_{n,l,i}^{(j)} \mathbf{h}_{n,k} + \sigma_c^2) \ln2}\Big) \ge LN \Gamma_\text{IR}, \\
		& \qquad \qquad \qquad \qquad \qquad \qquad \qquad \qquad \qquad \qquad \ \forall k\in\mathcal{K}_\text{IR}.
	\end{aligned}
\end{equation}
As a result, in each iteration \(j\), with given \(\{\mathbf{W}_{n,l,k}^{(j)}\}\), problem \eqref{1.1} is approximated by the following problem.
\begin{equation*}
	\begin{aligned} 
		\text{(P3):} \min_{\{\mathbf{W}_{n,l,k}\}, \zeta} &\ \mathcal{E}\big(\{\mathbf{W}_{n,l,k}\}, \zeta\big) \\
		\mathrm{s.t.} \quad &\ \eqref{SCA-C}, \ \eqref{1p}, \ \eqref{1b}, \ \text{and} \ \eqref{1sdp}.
	\end{aligned}
\end{equation*}
%whose optimal solution is updated as \(\{\mathbf{W}_{n,l,k}^{(j+1)}\}\) for the next iteration.
Notice that problem (P3) is convex, which can be efficiently solved via standard convex optimization tools such as CVX \cite{cvx2012}. Let \(\{\hat{\mathbf{W}}_{n,l,k}^{\text{SCA}}\}\) and \(\hat{\zeta}^{\text{SCA}}\) denote the obtained optimal solution to problem (P3), which is then used as the local point \(\{\mathbf{W}_{n,l,k}^{(j+1)}\}\) in the next iteration \(j+1\).

In summary, the SCA-based iterative algorithm for solving problem (P2) is presented as follows. First, we start by considering an initial local point \(\{\mathbf{W}_{n,l,k}^{(1)}\}\), which can be found via, e.g., the low-complexity heuristic designs in Section IV later. In each iteration \(j\) with local point \(\{\mathbf{W}_{n,l,k}^{(j)}\}\), we solve problem (P3) and accordingly update \(\{\mathbf{W}_{n,l,k}^{(j+1)}\}\) and \(\zeta\). The operation is iterated until a certain convergence criterion is met.
Notably, as (P3) consistently yields non-increasing objective values for (P2) over iterations, and the beampattern matching error in (P2) is lower bounded, the convergence of the SCA-based iterative algorithm is ensured.

\subsection{FP-based Algorithm for Solving (P2)}

In this subsection, we present an alternative iterative algorithm to solve problem (P2) via the technique of FP \cite{shen2018fractional, shen2018fractional2}. 
%We also employ FP for the complexity and convergence performance comparison with SCA. 
In this algorithm, we use the Lagrangian dual transform and the quadratic transform to deal with the non-convexity in \eqref{1c}.

%\subsubsection{Lagrangian Dual Transform}
To start with, we apply the Lagrangian dual transform to reformulate the constraints in \eqref{1c}. By introducing auxiliary variables \(\boldsymbol{\alpha} \triangleq \{\alpha_{n,l,k}\}\), \eqref{1c} is reformulated as \cite{shen2018fractional2}
\begin{equation} \label{LDT-C}
	\begin{aligned}
		&\frac{1}{LN} \sum_{l\in\mathcal{L}} \sum_{n\in\mathcal{N}} \Big( \log_2(1 + \alpha_{n,l,k}) - \alpha_{n,l,k} + (1+\alpha_{n,l,k}) \\ & \cdot \frac{\mathbf{h}_{n,k}^H \mathbf{W}_{n,l,k} \mathbf{h}_{n,k}}{\sum_{i\in\{0\}\cup\mathcal{K}_\text{IR}} \mathbf{h}_{n,k}^H \mathbf{W}_{n,l,i} \mathbf{h}_{n,k} + \sigma_c^2} \Big) \ge \Gamma_\text{IR}, \forall k\in\mathcal{K}_\text{IR}.
	\end{aligned}
\end{equation}
Consequently, problem (P2) is reformulated as
\begin{equation*}
	\begin{aligned} 
		\text{(P4.1):} \min_{\{\mathbf{W}_{n,l,k}\}, \zeta, \boldsymbol{\alpha}} &\ \mathcal{E}\big(\{\mathbf{W}_{n,l,k}\}, \zeta\big) \\
		\mathrm{s.t.} \quad \ &\ \eqref{LDT-C}, \ \eqref{1p}, \ \eqref{1b}, \ \text{and} \ \eqref{1sdp}.
	\end{aligned}
\end{equation*}
To solve problem (P4.1), we alternately optimize \(\{\mathbf{W}_{n,l,k}\}\), \(\zeta\), and \(\boldsymbol{\alpha}\), with the others being given. 

First, given \(\{\mathbf{W}_{n,l,k}\}\) and \(\zeta\), the optimal \(\boldsymbol{\alpha}\) for problem (P4.1) can be explicitly determined as follows by setting the derivative of the left-hand side (LHS) of \eqref{LDT-C} to be zero, \(\forall n\in\mathcal{N}, l\in\mathcal{L}, k\in\mathcal{K}_\text{IR}\).
\begin{equation}\label{alpha-star}
	\alpha_{n,l,k}^\star = \frac{\mathbf{h}_{n,k}^H \mathbf{W}_{n,l,k} \mathbf{h}_{n,k}}{\sum_{i\in\{0\}\cup\mathcal{K}_\text{IR}\setminus\{k\}} \mathbf{h}_{n,k}^H \mathbf{W}_{n,l,i} \mathbf{h}_{n,k} + \sigma_c^2}.
\end{equation}

Next, we optimize \(\{\mathbf{W}_{n,l,k}\}\) and \(\zeta\) in problem (P4.1) with given \(\boldsymbol{\alpha}\) by applying the quadratic transform on the fractional terms in \eqref{LDT-C}.
By introducing auxiliary variables \(\boldsymbol{\beta} \triangleq \{\beta_{n,l,k}\}\), \eqref{LDT-C} is reformulated as \cite{shen2018fractional2}
\begin{equation} \label{QT-C}
	\begin{aligned}
		&\frac{1}{LN} \sum_{l\in\mathcal{L}} \sum_{n\in\mathcal{N}} \Big( \log_2(1 + \alpha_{n,l,k}) - \alpha_{n,l,k} \\
		&+ 2\sqrt{1+\alpha_{n,l,k}} \beta_{n,l,k} \sqrt{\mathbf{h}_{n,k}^H \mathbf{W}_{n,l,k} \mathbf{h}_{n,k}} \\ &- \beta_{n,l,k}^2 \big(\sum_{i\in\{0\}\cup\mathcal{K}_\text{IR}} \mathbf{h}_{n,k}^H \mathbf{W}_{n,l,i} \mathbf{h}_{n,k} + \sigma_c^2\big) \Big) \ge  \Gamma_\text{IR}, \forall k\in\mathcal{K}_\text{IR},
	\end{aligned}
\end{equation}
which is convex w.r.t. \(\{\mathbf{W}_{n,l,k}\}\). Therefore, in order to solve problem (P4.1) over \(\{\mathbf{W}_{n,l,k}\}\) and \(\zeta\) with fixed \(\boldsymbol{\alpha}\), we can equivalently solve the following problem over \(\{\mathbf{W}_{n,l,k}\}\), \(\zeta\), and \(\boldsymbol{\beta}\).
\begin{equation*}
	\begin{aligned} 
		\text{(P4.2):} \min_{\{\mathbf{w}_{n,l,k}\}, \zeta, \boldsymbol{\beta}} &\ \mathcal{E}\big(\{\mathbf{W}_{n,l,k}\}, \zeta\big) \\
		\mathrm{s.t.} \quad \ &\ \eqref{QT-C}, \ \eqref{1p}, \ \eqref{1b}, \ \text{and} \ \eqref{1sdp}.
	\end{aligned}
\end{equation*}
%The overall optimization strategy to problem (P2) is then to alternatively optimize \(\boldsymbol{\alpha}\) according to \eqref{alpha-star} and optimize \(\boldsymbol{\beta}\) and \(\big(\{\mathbf{W}_{n,l,k}\}, \zeta\big)\) as in \eqref{QT-P}. 
Specifically, problem (P4.2) is solved via alternately optimizing \(\{\mathbf{W}_{n,l,k}\}\), \(\zeta\), and \(\boldsymbol{\beta}\). With given \(\{\mathbf{W}_{n,l,k}\}\) and \(\zeta\), the optimal \(\boldsymbol{\beta}\) can be found in the following by setting the derivative of the LHS of \eqref{QT-C} to be zero, \(\forall n\in\mathcal{N}, l\in\mathcal{L}, k\in\mathcal{K}_\text{IR}\).
\begin{equation}\label{beta-star}
	\beta_{n,l,k}^\star = \frac{\sqrt{1+\alpha_{n,l,k}} \sqrt{\mathbf{h}_{n,k}^H \mathbf{W}_{n,l,k} \mathbf{h}_{n,k}}}{\sum_{i\in\{0\}\cup\mathcal{K}_\text{IR}} \mathbf{h}_{n,k}^H \mathbf{W}_{n,l,i} \mathbf{h}_{n,k} + \sigma_c^2}.
\end{equation}
With given \(\boldsymbol{\beta}\), the optimization of \(\big(\{\mathbf{W}_{n,l,k}\}, \zeta\big)\) in problem (P4.2) is expressed as
\begin{equation*}
	\begin{aligned} 
		\text{(P4.3):} \min_{\{\mathbf{W}_{n,l,k}\}, \zeta} &\ \mathcal{E}\big(\{\mathbf{W}_{n,l,k}\}, \zeta\big) \\
		\mathrm{s.t.} \quad &\ \eqref{QT-C}, \ \eqref{1p}, \ \eqref{1b}, \ \text{and} \ \eqref{1sdp}.
	\end{aligned}
\end{equation*}
Notice that problem (P4.3) is convex, allowing us to use standard convex optimization tools such as CVX \cite{cvx2012} to efficiently find its optimal solution, denoted by \(\{\hat{\mathbf{W}}_{n,l,k}^{\text{FP}}\}\) and \(\hat{\zeta}^{\text{FP}}\). Therefore, problems (P4.3) and accordingly (P4.1) are finally solved.

To conclude, the FP-based iterative algorithm for solving problem (P2) is presented as follows. First, we find an initial point of \(\{\mathbf{W}_{n,l,k}\}\) (e.g., via the low-complexity heuristic designs in Section IV), together with those for \(\boldsymbol{\alpha}\) and \(\boldsymbol{\beta}\). In each iteration, \(\boldsymbol{\alpha}\), \(\boldsymbol{\beta}\), as well as \(\{\mathbf{W}_{n,l,k}\}\) and \(\zeta\) are alternately updated in \eqref{alpha-star}, \eqref{beta-star}, and by solving problem (P4.3), respectively, with the others given. The operation is iterated until a convergence criterion is met. Notably, as the iterations consistently yield non-increasing objective values for (P2), and the beampattern matching error in (P2) is lower bounded, the convergence of the FP-based iterative algorithm is ensured.
%As the updated beampattern matching error consistently yields non-increasing over iterations, and the objective of (P2) is lower bounded, the convergence of the algorithm is assured.

It is worth comparing the SCA and FP-based algorithms for solving problem (P2). It is shown that both algorithms achieve converged solutions, and it will be shown in Section V that they will achieve similar performance but with different complexities depending on the specific CVX solvers (e.g., SDPT3, SeDuMi, and Mosek) employed for solving problems (P3) and (P4.3), respectively. 

\section{Other Heuristic Solutions}

In this section, we propose three heuristic algorithms to solve the considered beampattern matching error minimization problem based on ZF beamforming, round-robin user scheduling, and time switching design principles, respectively, which generally have lower implementation complexity and will serve as benchmarks for performance comparison later.

\subsection{ZF Based Beamforming Design}

Based on the ZF principle, we first design the information beamforming to avoid inter-user interference at IRs. 
%caused by the information-bearing signal to other IRs and the dedicated sensing/powering signal.
To be specific, at each OFDM symbol \(l\in\mathcal{L}\) and subcarrier \(n\in\mathcal{N}\), we ensure that the information beamforming vector \(\mathbf{w}_{n,l,k}\) for each IR \(k\in\mathcal{K}_\text{IR}\) is orthogonal to the communication channels of other IRs, i.e., \(|\mathbf{h}_{n,k}^H \mathbf{w}_{n,l,i}|^2 = 0\), \(\forall k,i \in\mathcal{K}_\text{IR}, k\ne i\). We set
\vspace{-0.1cm}\begin{equation}
	\mathbf{w}_{n,l,k} = \sqrt{\tilde{p}_{n,l,k}} \tilde{\mathbf{w}}_{n,k},
\vspace{-0.1cm}\end{equation}
where \(\tilde{p}_{n,l,k}\) denotes the transmit power of the associated beam to be optimized, and the unit vector \(\tilde{\mathbf{w}}_{n,k}\) is obtained as
\vspace{-0.1cm}\begin{equation}
	\tilde{\mathbf{w}}_{n,k} = \frac{[\mathbf{H}_{n}^{\dag}]_{:,k}}{\|[\mathbf{H}_{n}^{\dag}]_{:,k}\|}.
\vspace{-0.1cm}\end{equation}
Here, \(\mathbf{H}_{n} = [\mathbf{h}_{n,1}, \dots, \mathbf{h}_{n,K_\text{IR}}]\) and \(\mathbf{H}_{n}^{\dag} = \mathbf{H}_{n} (\mathbf{H}_{n}^H \mathbf{H}_{n})^{-1}\) denote the communication channel matrix to all \(K_\text{IR}\) IRs and its pseudo-inverse, respectively.

Next, we design the sensing/energy beamforming vectors, which lie in the null space of the communication channels to avoid the co-channel interference towards IRs, i.e., \(\mathbf{h}_{n,k}^H \mathbf{W}_{n,l,0} \mathbf{h}_{n,k} = 0, \forall n\in\mathcal{N}, l\in\mathcal{L}, k\in\mathcal{K}_\text{IR}\).
At each OFDM symbol \(l\) and subcarrier \(n\), we construct the covariance matrix of the dedicated sensing/energy signal as
\vspace{-0.1cm}\begin{equation}
	\mathbf{W}_{n,l,0} = \mathbf{U}_{n,0} \tilde{\mathbf{P}}_{n,l,0} \mathbf{U}_{n,0}^H,
\vspace{-0.1cm}\end{equation}
where \(\tilde{\mathbf{P}}_{n,l,0} \in \mathbb{C}^{(N_t-K_\text{IR})\times (N_t-K_\text{IR})}\) is a positive semidefinite matrix to be optimized. Here, the eigenvalue decomposition (EVD) of \(\mathbf{H}_n \mathbf{H}_n^H\) is defined as
\vspace{-0.1cm}\begin{equation}
	\mathbf{H}_n \mathbf{H}_n^H = \begin{bmatrix}
		\mathbf{U}_{n,1} & \mathbf{U}_{n,0}
	\end{bmatrix} \begin{bmatrix}
		\mathbf{\Lambda}_{n,1} & \mathbf{0} \\ \mathbf{0} & \mathbf{0}
	\end{bmatrix} \begin{bmatrix}
		\mathbf{U}_{n,1}^H \\ \mathbf{U}_{n,0}^H
	\end{bmatrix},
\vspace{-0.1cm}\end{equation}
where \(\mathbf{U}_{n,1} \in \mathbb{C}^{N_t\times K_\text{IR}}\) and \(\mathbf{U}_{n,0} \in \mathbb{C}^{N_t\times (N_t-K_\text{IR})}\) contain the eigenvectors corresponding to the \(K_\text{IR}\) non-zero eigenvalues and the \(N_t-K_\text{IR}\) zero eigenvalues of \(\mathbf{H}_n \mathbf{H}_n^H\), respectively.

Therefore, based on the above ZF-based beamforming design, problem (P1) is reduced to the following optimization problem. Here, for ease of notation, we define \(\lambda_{n,k} = |\mathbf{h}_{n,k}^H \tilde{\mathbf{w}}_{n,k}|\), \(\delta_{m,n,k} = |\mathbf{v}^T(\theta_m) \tilde{\mathbf{w}}_{n,k}|\), \(\boldsymbol{\delta}_{m,n,0}^H =\mathbf{v}^T(\theta_m) \mathbf{U}_{n,0}\), \(\rho_{n,i,k} = |\mathbf{g}_{n,i}^H \tilde{\mathbf{w}}_{n,k}|\), and \(\boldsymbol{\rho}_{n,i,0}^H = \mathbf{g}_{n,i}^H \mathbf{U}_{n,0}\).
\vspace{-0.1cm}\begin{equation*} \label{59}
	\begin{aligned} 
		\text{(P5):} &\min_{\{\tilde{p}_{n,l,k}\}, \{\tilde{\mathbf{P}}_{n,l,0} \succeq \mathbf{0}\}, \zeta} \ \sum_{l\in\mathcal{L}} \sum_{m\in\mathcal{M}} \Big|\sum_{n\in\mathcal{N}} \big(\sum_{k\in\mathcal{K}_\text{IR}} \delta_{m,n,k}^2 \tilde{p}_{n,l,k} \\ &\qquad \qquad \qquad \qquad \ + \boldsymbol{\delta}_{m,n,0}^H \tilde{\mathbf{P}}_{n,l,0} \boldsymbol{\delta}_{m,n,0}\big) - \zeta \mathcal{P}_l(\theta_m)\Big|^2 \\
		\mathrm{s.t.} &\ \frac{1}{LN} \sum_{l\in\mathcal{L}} \sum_{n\in\mathcal{N}} \log_2\Big(1 + \frac{\lambda_{n,k}^2 \tilde{p}_{n,l,k}}{\sigma_c^2}\Big) \ge \Gamma_\text{IR}, \forall k \in \mathcal{K}_\text{IR}, \\
		&\ \frac{1}{L} \sum_{l\in\mathcal{L}} \sum_{n\in\mathcal{N}} \big(\sum_{k\in\mathcal{K}_\text{IR}} \rho_{n,i,k}^2 \tilde{p}_{n,l,k} + \boldsymbol{\rho}_{n,i,0}^H \tilde{\mathbf{P}}_{n,l,0} \boldsymbol{\rho}_{n,i,0}\big) \\
		&\qquad \qquad \qquad \qquad \qquad \qquad \quad \ \ \ge \Gamma_\text{ER}, \forall i \in \mathcal{K}_\text{ER}, \\ 
		&\ \sum_{n\in\mathcal{N}} \big(\sum_{k\in\mathcal{K}_\text{IR}} \tilde{p}_{n,l,k} + \mathrm{tr}(\tilde{\mathbf{P}}_{n,l,0})\big) = P_0, \forall l\in\mathcal{L}.
	\end{aligned}
\vspace{-0.1cm}\end{equation*}
Notice that problem (P5) is convex and can be directly solved by standard convex optimization tools. 
%The time complexity for solving (P5) is significantly lower compared to (P1),
Compared with problem (P1), problem (P5) does not require any iteration like SCA and FP-based algorithms, thus having lower implementation complexity.

\subsection{Round-robin User Scheduling}

In this scheme, the BS communicates with each IR \(k\in\mathcal{K}_\text{IR}\) in a round-robin manner. Towards this end, the transmitted signal at each OFDM symbol \(l\in\mathcal{L}\) and subcarrier \(n\in\mathcal{N}\) is designed as
\vspace{-0.1cm}\begin{equation}
	\mathbf{x}_{n,l} = \mathbf{w}_{n,l,k_{n,l}} s_{n,l,k_{n,l}} + \mathbf{s}_{n,l,0},
\vspace{-0.1cm}\end{equation}
%which contains the information signal to IR \(k_{n,l} \triangleq \{k|\mathrm{mod}(k,K_\text{IR}) = \mathrm{mod}(n+(l-1)N,K_\text{IR})\}\) together with the dedicated sensing/powering signal.
i.e., at each symbol \(l\) and subcarrier \(n\), only one IR \(k_{n,l} \triangleq \{k|\mathrm{mod}(k,K_\text{IR}) = \mathrm{mod}(n+(l-1)N,K_\text{IR})\}\) is scheduled for communication. In other words, at each symbol \(l\), each IR \(k\) is scheduled only at subcarriers \(n \in \mathcal{N}_{l,k} \triangleq \{n|\mathrm{mod}(n+(l-1)N,K_\text{IR}) = \mathrm{mod}(k,K_\text{IR})\}\).
Additionally, at each symbol \(l\), the transmit power is equally allocated among all \(N\) subcarriers, with the transmit power constraint at the BS given by \(\mathbb{E}[\mathbf{x}_{n,l} \mathbf{x}_{n,l}^H] = \mathbf{W}_{n,l,k_{n,l}} + \mathbf{W}_{n,l,0} = \frac{P_0}{N}\), \(\forall n\in\mathcal{N}, l\in\mathcal{L}\).
In this case, the corresponding beampattern matching error minimization problem is given by
\vspace{-0.1cm}\begin{equation*} \label{61}
	\begin{aligned} 
		\text{(P6):} &\min_{\{\mathbf{W}_{n,l,k}\}, \zeta} \ \sum_{l\in\mathcal{L}} \sum_{m\in\mathcal{M}} \Big|\sum_{n\in\mathcal{N}} \mathbf{v}^T(\theta_m) \\
		&\qquad \qquad \ \ \cdot(\mathbf{W}_{n,l,k_{n,l}} + \mathbf{W}_{n,l,0}) \mathbf{v}^*(\theta_m) - \zeta \mathcal{P}_l(\theta_m)\Big|^2 \\
		\mathrm{s.t.} &\ \frac{1}{LN} \sum_{l\in\mathcal{L}} \sum_{n\in\mathcal{N}_{l,k}} \log_2\Big(1 + \frac{\mathbf{h}_{n,k}^H \mathbf{W}_{n,l,k} \mathbf{h}_{n,k}}{\mathbf{h}_{n,k}^H \mathbf{W}_{n,l,0} \mathbf{h}_{n,k} + \sigma_c^2}\Big) \\ 
		&\qquad \qquad \qquad \qquad \qquad \qquad \quad \ge \Gamma_\text{IR}, \forall k\in\mathcal{K}_\text{IR}, \\
		&\ \frac{1}{L} \sum_{l\in\mathcal{L}} \sum_{n\in\mathcal{N}} \mathbf{g}_{n,i}^H (\mathbf{W}_{n,l,k_{n,l}} + \mathbf{W}_{n,l,0}) \mathbf{g}_{n,i}  \\ 
		&\ \qquad \qquad \qquad \qquad \quad \ \ge \Gamma_\text{ER}, \forall i \in \mathcal{K}_\text{ER}, \\ 
		&\ \mathrm{tr}(\mathbf{W}_{n,l,k_{n,l}} + \mathbf{W}_{n,l,0}) = \frac{P_0}{N}, \forall n\in\mathcal{N}, l\in\mathcal{L}, \\
		&\ \eqref{1sdp} \ \text{and} \ \eqref{1one}.
	\end{aligned}
\vspace{-0.1cm}\end{equation*}
Similar as in Proposition 1, it can be shown that the SDR of problem (P6) is tight, i.e., by dropping the rank-one constraint \eqref{1one} in problem (P6), the resultant problem (SDR6) can be shown to have the same optimal objective value. Furthermore, it is easy to verify that at the optimality of the relaxed problem, it holds that \(\{\mathbf{W}_{n,l,0} = \mathbf{0}\}\), such that problem (SDR6) is reduced to a convex problem that can be optimally solved by standard convex optimization tools. Compared with problem (P1), no iterations based on SCA or FP are needed for problem (P6), thus leading to lower complexity again.

\subsection{Time-switching Design}

In this scheme, the sensing, communication, and powering oriented beamforming designs are implemented in a time-switching manner for performance comparison. To be specific, the transmission duration is divided into three orthogonal phases with portions \(\{t_1,t_2,t_3\}\), where \(t_1 + t_2 + t_3 = 1\).\footnote{It is assumed that the transmission duration can be arbitrarily partitioned, such that the time portions \(t_j\)'s are viewed as continuous variables. This consideration is adopted to characterize the performance for comparison only.} Here, each phase is dedicated to one function of sensing, communication, and powering, respectively. As such, the corresponding design in each phase \(j \in \{1,2,3\}\) is formulated as the following problem (P7.\(j\)).
\vspace{-0.1cm}\begin{equation*}
	\begin{aligned} 
		\text{(P7.1):} \min_{\{\mathbf{W}_{n,l,0}\}, \zeta} \ \mathcal{E}\big(\{\mathbf{W}_{n,l,k}\}, \zeta\big), \
		\mathrm{s.t.} \ \eqref{1b} \ \text{and} \ \eqref{1sdp},
	\end{aligned}
\vspace{-0.1cm}\end{equation*}
\vspace{-0.1cm}\begin{equation*}
	\begin{aligned} 
		\text{(P7.2):} \max_{\{\mathbf{W}_{n,l,k}\}_{k\in\mathcal{K}_\text{IR}}} \min_{k\in\mathcal{K}_\text{IR}} &\ R_{k}\big(\{\mathbf{W}_{n,l,k}\}\big) \\
		\mathrm{s.t.} &\ \eqref{1b}, \ \eqref{1sdp}, \ \text{and} \ \eqref{1one},
	\end{aligned}
\vspace{-0.1cm}\end{equation*}
\vspace{-0.1cm}\begin{equation*}
	\begin{aligned} 
		\text{(P7.3):} \max_{\{\mathbf{W}_{n,l,0}\}} \min_{i\in\mathcal{K}_\text{ER}} \ P_{i}^\text{ER}\big(\{\mathbf{W}_{n,l,k}\}\big), \
		\mathrm{s.t.} \ \eqref{1b} \ \text{and} \ \eqref{1sdp}.
	\end{aligned}
\vspace{-0.1cm}\end{equation*}
In problems (P7.1) and (P7.3), only the dedicated sensing/energy signals are considered for beampattern matching and harvested power maximization, respectively, with \(\{\mathbf{W}_{n,l,k} = \mathbf{0}\}_{k\in\mathcal{K}_\text{IR}}\). Therefore, (P7.1) and (P7.3) are convex and can be solved by convex optimization tools. Conversely, problem (P7.2) only considers the information signals for communication rate maximization with \(\{\mathbf{W}_{n,l,0} = \mathbf{0}\}\), which is non-convex and can be solved similarly as problem (P1) via SDR and SCA/FP-based algorithms. We denote the obtained solution of \(\{\mathbf{W}_{n,l,k}\}\) for (P7.\(j\)) as \(\{\check{\mathbf{W}}_{n,l,k}^{(j)}\}\). The corresponding achieved beampattern gain, communication rate, and harvested power by each problem (P7.\(j\)) are given as follows, respectively.
\vspace{-0.0cm}\begin{subequations}
	\begin{align} 
		\check{\mathcal{G}}_{m,l}^{(j)} &= \mathcal{G}_l\big(\theta_m, \{\check{\mathbf{W}}_{n,l,k}^{(j)}\}\big), \forall m \in \mathcal{M}, l \in \mathcal{L}, \\
		\check{R}_k^{(j)} &= R_k\big(\{\check{\mathbf{W}}_{n,l,k}^{(j)}\}\big), \forall k \in \mathcal{K}_\text{IR}, \\
		\check{P}_i^{(j)} &= P_{i}^\text{ER}\big(\{\check{\mathbf{W}}_{n,l,k}^{(j)}\}\big), \forall i \in \mathcal{K}_\text{ER}.
	\end{align}
\vspace{-0.1cm}\end{subequations}

During each phase \(j\in\{1,2,3\}\), the BS employs the dedicated transmission design \(\{\check{\mathbf{W}}_{n,l,k}^{(j)}\}\) for sensing, communication, and powering, respectively.
Then, we optimize the time allocation among the three phases, for which the corresponding beampattern matching error minimization problem is given by
\vspace{-0.1cm}\begin{equation*} \label{63}
	\begin{aligned} 
		\text{(P8):} \min_{\{t_j \ge 0\}, \zeta} &\ \sum_{l\in\mathcal{L}} \sum_{m\in\mathcal{M}} \Big|\sum_{j\in\{1,2,3\}} t_j \check{\mathcal{G}}_{m,l}^{(j)} - \zeta \mathcal{P}_l(\theta_m)\Big|^2 \\
		\mathrm{s.t.} \ \ &\ t_2 \check{R}_k^{(2)} \ge \Gamma_\text{IR}, \forall k \in \mathcal{K}_\text{IR}, \\
		&\ t_1 \check{P}_i^{(1)} + t_2 \check{P}_i^{(2)} + t_3 \check{P}_i^{(3)} \ge \Gamma_\text{ER}, \forall i \in \mathcal{K}_\text{ER}, \\
		&\ t_1 + t_2 + t_3 = 1,
	\end{aligned}
\vspace{-0.1cm}\end{equation*}
where the sensing is implemented by collecting all data over the three phases, and the ERs can harvest power from both information and sensing/energy signals.
Problem (P8) is convex and can be solved by using standard convex optimization tools.

\section{Numerical Results}

This section provides numerical results to evaluate the performance of our proposed designs. The following system parameters are considered unless specified otherwise.
The BS is equipped with a ULA of \(N_t = 16\) transmit antennas and \(N_r = 32\) receive antennas with half-wavelength spacing between adjacent antennas. 
We consider the OFDM transmission over \(L = 256\) symbols, each consisting of \(N = 16\) subcarriers and \(N_\text{CP} = 4\) samples for cyclic prefix (CP). In addition, the carrier frequency is set to \(f_c = 28\) GHz. The subcarrier spacing is \(\Delta_f = 120\) KHz, the total bandwidth is \(B = 1.92\) MHz, and the transmission duration for each symbol is \(T_\text{sym} = 8.333\) \(\mu\)s.
%\(T_\text{sym} = 10.417\) \(\mu\)s.

\begin{figure}[tb]
	\centering {\includegraphics[width=0.35\textwidth]{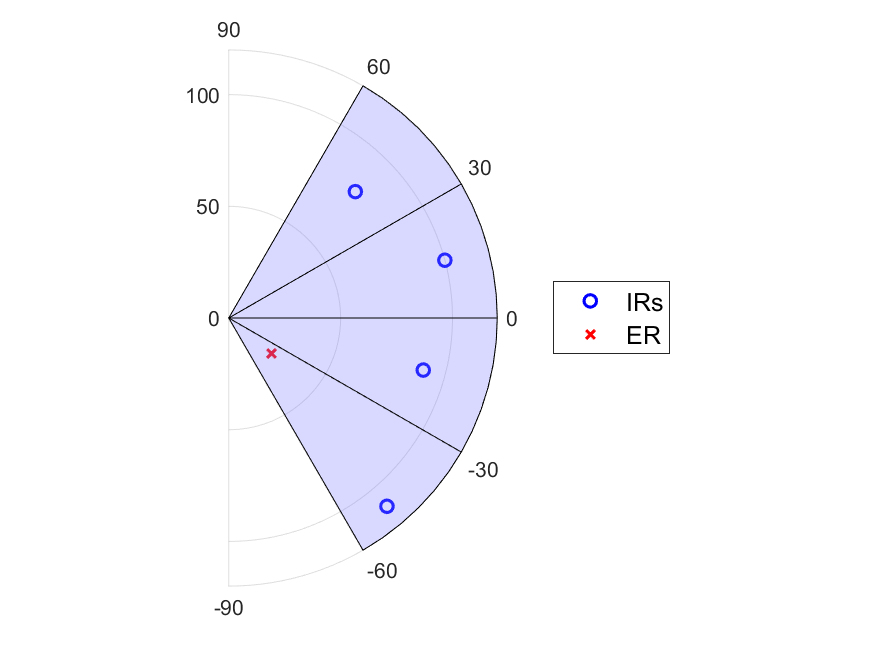}} 
	\caption{Illustration of interested sensing areas over time slots and the locations of IRs and the ER.}
	\label{Fig_setup}
\vspace{-0.1cm}\end{figure}
As depicted in Fig. \ref{Fig_setup}, during beam scanning, the BS senses angles within the range of \([-60^\circ, 60^\circ]\). We partition the \(256\) OFDM symbols into \(Q = 4\) time slots, for which the desired beampatterns are centered on angles \(\{-45^\circ, -15^\circ, 15^\circ, 45^\circ\}\), respectively, with a width of \(30^\circ\).\footnote{For illustration, we assume non-overlapping interested angles per slot.} There are \(K_\text{IR} = 4\) IRs positioned at angles \(\{-50^\circ, -15^\circ, 15^\circ, 45^\circ\}\) and distances \(\{110, 90, 100, 80\}\) m, respectively, together with an ER positioned at \(-40^\circ\) and \(25\) m. The angular grids are specified by \(M = 48\) discrete angles spanning uniformly over \([-\frac{\pi}{2}, \frac{\pi}{2}]\). In addition, the demonstrated beampattern gain and matching error are normalized by \(\frac{1}{\zeta}\) and \(\frac{1}{L M \zeta^2}\), respectively.

We consider the Rician fading channels for both communication and WPT, where the Rician factor is set as \(\kappa = 20\). The path loss is given by \(L_p = K_{\text{ref}}(\frac{D}{D_{\text{ref}}})^{\eta_p}\), where \(D\) represents the distance between the BS and each IR/ER, \(D_{\text{ref}} = 1\) m is the reference distance, \(K_{\text{ref}} = 30\) dB is the path loss at \(D_{\text{ref}}\), and \(\eta_p = 3\) is the path loss exponent. Additionally, the transmit power at the BS is set as \(P_0 = 30\) dBm, and the noise power at the BS receiver and IRs is set as \(-70\) dBm.

\begin{figure}[tb]
	\centering 
	\subfigure[Beampattern matching error over iterations.]
	{\includegraphics[width=0.8\columnwidth]{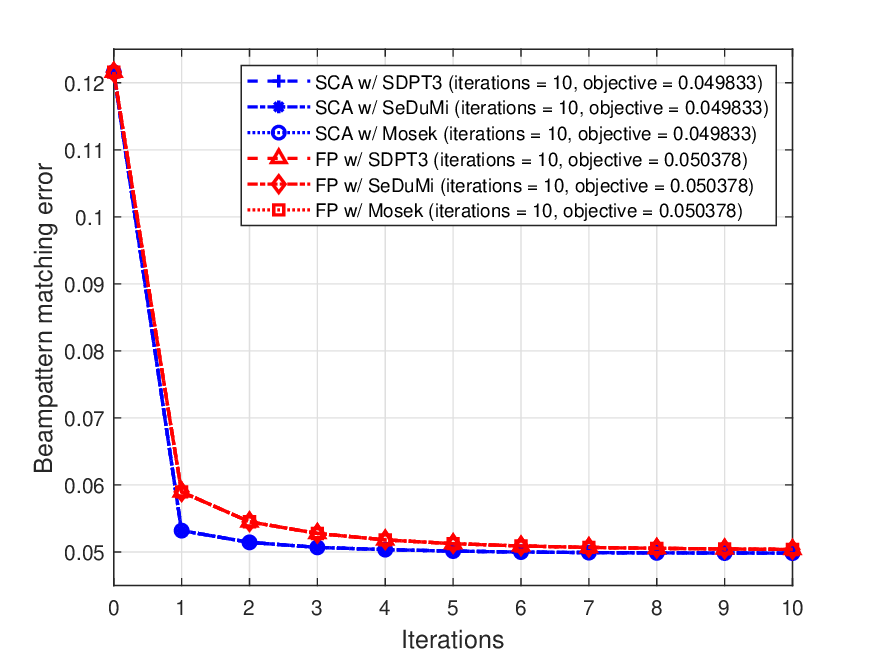}}
	\subfigure[Execution time over iterations.]
	{\includegraphics[width=0.8\columnwidth]{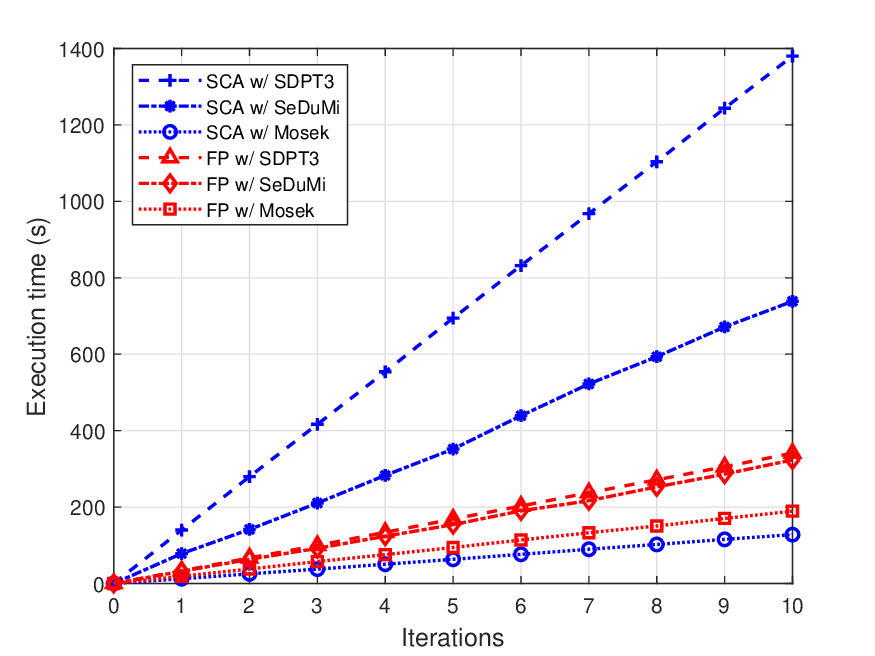}}
	\caption{Convergence and complexity performance for SCA and FP algorithms under different CVX solvers with \(\Gamma_\text{IR} = 200\) Kbps and \(\Gamma_\text{ER} = 10\) \(\mu\)W.}
	\label{Fig_time}
\vspace{-0.1cm}\end{figure}
Fig. \ref{Fig_time} shows the convergence and complexity performance for the SCA and FP-based algorithms under different CVX solvers (SDPT3, SeDuMi, and Mosek), with \(\Gamma_\text{IR} = 200\) Kbps and \(\Gamma_\text{ER} = 10\) \(\mu\)W. In the algorithms for solving (P2), the solution of ZF beamforming is utilized as the initial point and the prescribed converge accuracy is set as \(10^{-3}\). The results include the objective value and execution time for the first \(10\) iterations, in which the algorithms are implemented based on Intel\textsuperscript{\textregistered} Core\textsuperscript{\texttrademark} i7-12700 Processor. 
As shown in Fig. \ref{Fig_time}(a), it is observed that both SCA and FP algorithms achieve non-increasing objectives, converging within the prescribed accuracy. The SCA algorithm demonstrates faster convergence compared to the FP algorithm from the given initial point, and requires fewer iterations to reach the convergence threshold.
In addition, it is observed in Fig. \ref{Fig_time}(b) that the execution time for both SCA and FP algorithms is significantly influenced by the choice of CVX solvers.
To be specific, the FP algorithm requires much less execution time than the SCA algorithm when using SDPT3 and SeDuMi solvers, while the Mosek solver greatly accelerates the SCA algorithm and makes it faster than the FP algorithm.
This improvement is due to Mosek's native support for handling models with logarithmic functions such as problem (P3) with the sum-logarithm constraint \eqref{SCA-C}, which is processed more efficiently than the experimental successive approximation method used by SDPT3 and SeDuMi \cite{cvx2012}. By contrast, the three solvers have comparable capacities in handling models like problem (P4.3) with the sum-square-root constraint \eqref{QT-C}.

\begin{figure*}[htb]
	\centering 
	\subfigure[Beampattern matching at each slot.] {\includegraphics[width=0.64\columnwidth]{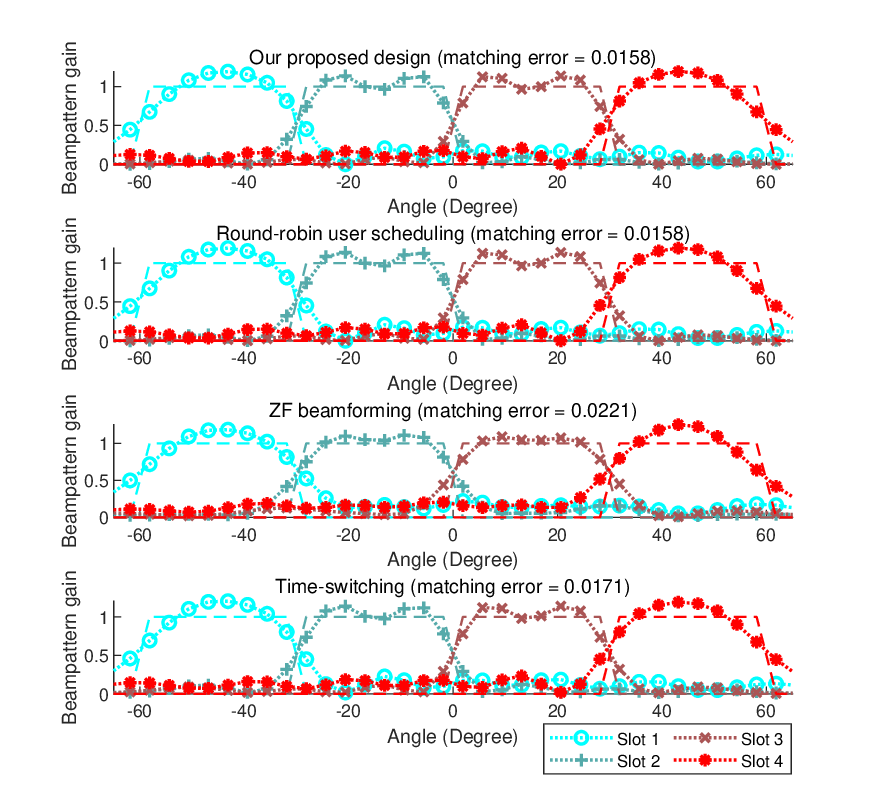}}
	\subfigure[Communication rate of each IR at each slot.] {\includegraphics[width=0.68\columnwidth]{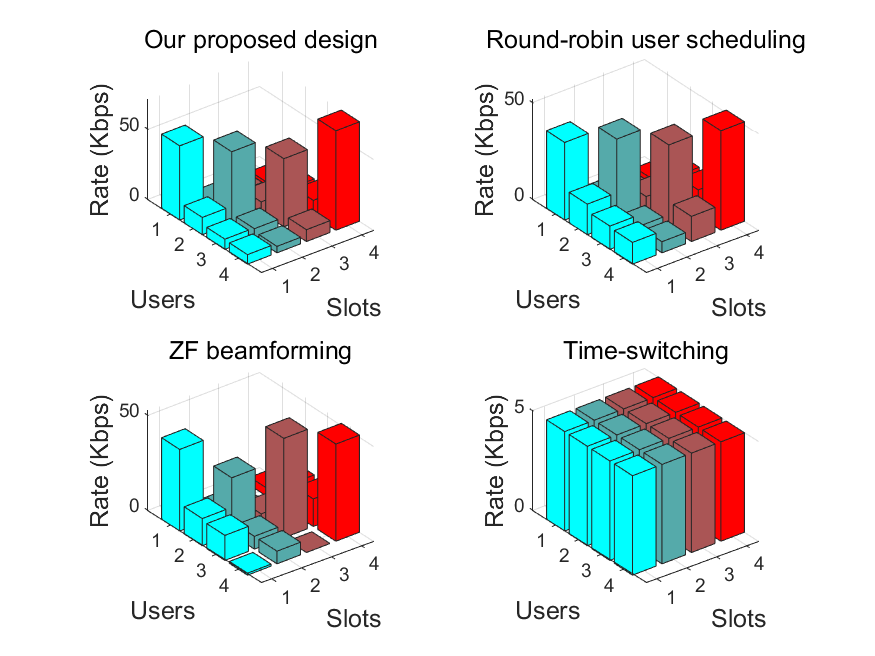}}
	\subfigure[Harvested power of the ER at each slot.] {\includegraphics[width=0.68\columnwidth]{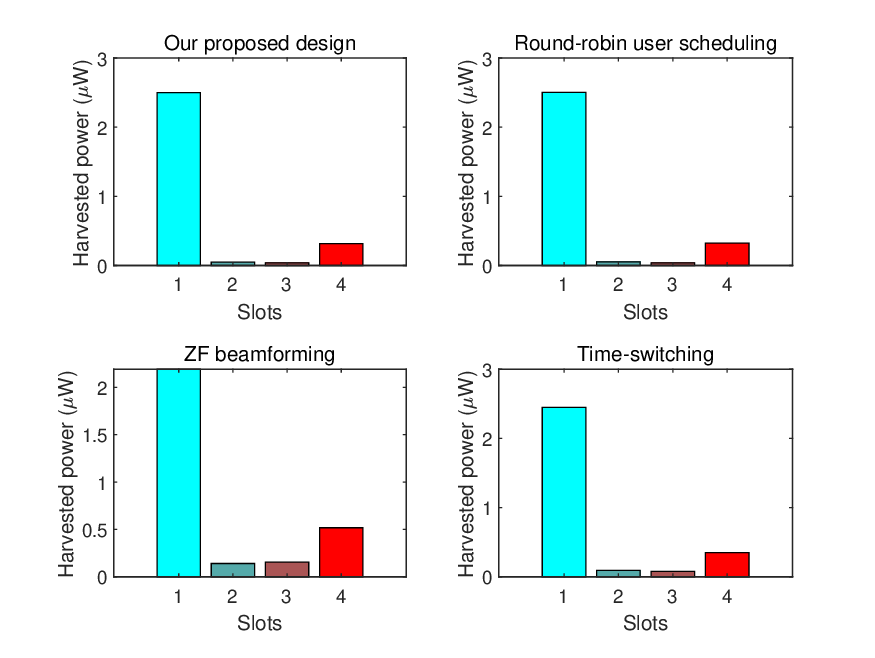}}
	\caption{Beampattern matching performance and the resultant communication rates at IRs and harvested power at the ER, where low communication rate constraint \(\Gamma_\text{IR} = 20\) Kbps and low harvested power constraint \(\Gamma_\text{ER} = 1\) \(\mu\)W are considered.}
	\label{Fig_LL}
\vspace{-0.1cm}\end{figure*}
\begin{figure*}[htb]
	\centering 
	\subfigure[Beampattern matching at each slot.] {\includegraphics[width=0.64\columnwidth]{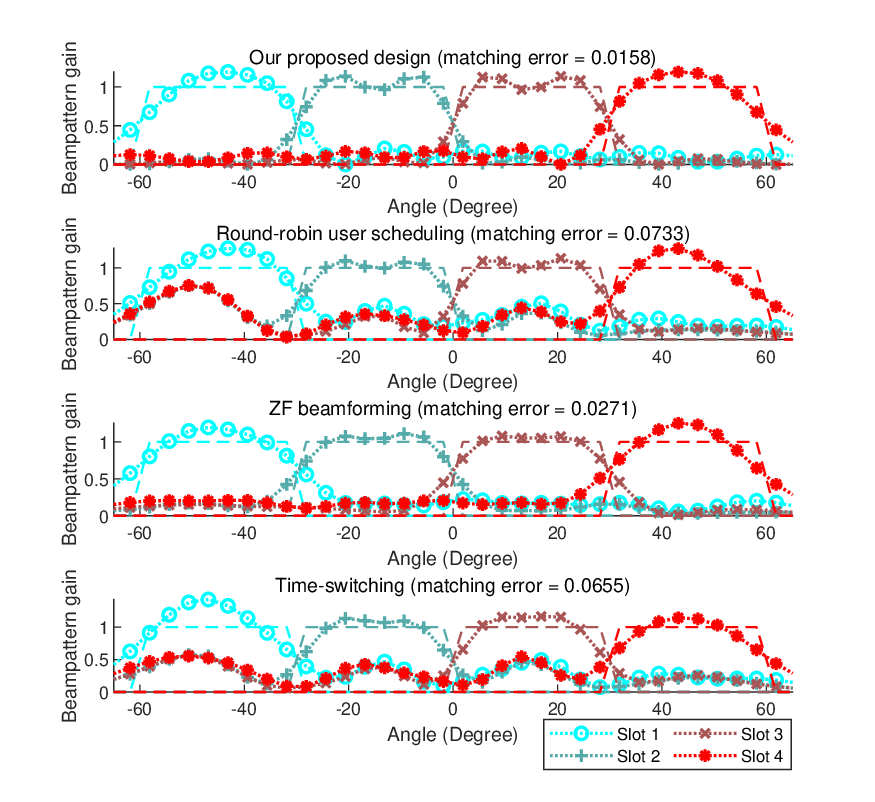}}
	\subfigure[Communication rate of each IR at each slot.] {\includegraphics[width=0.68\columnwidth]{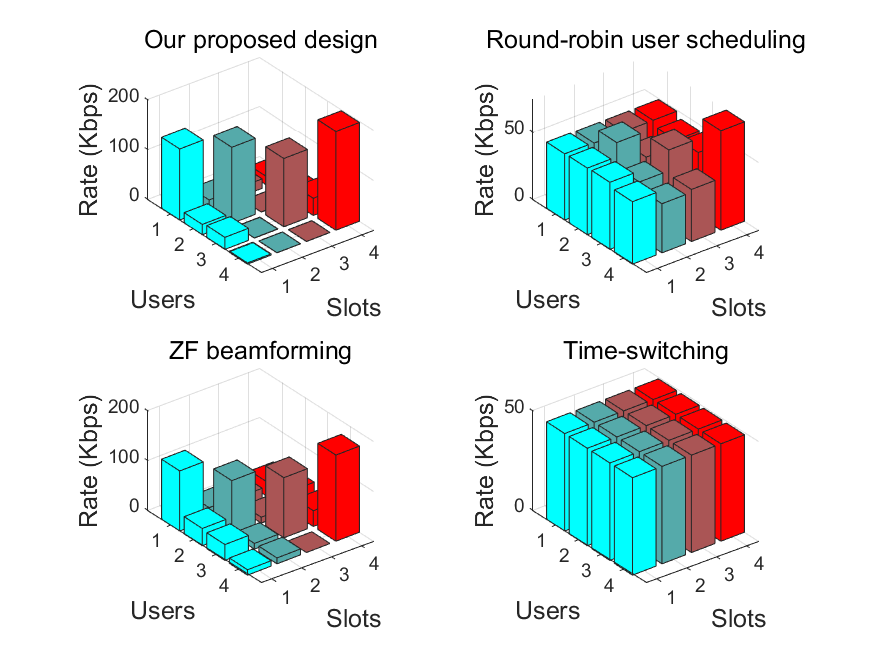}}
	\subfigure[Harvested power of the ER at each slot.] {\includegraphics[width=0.68\columnwidth]{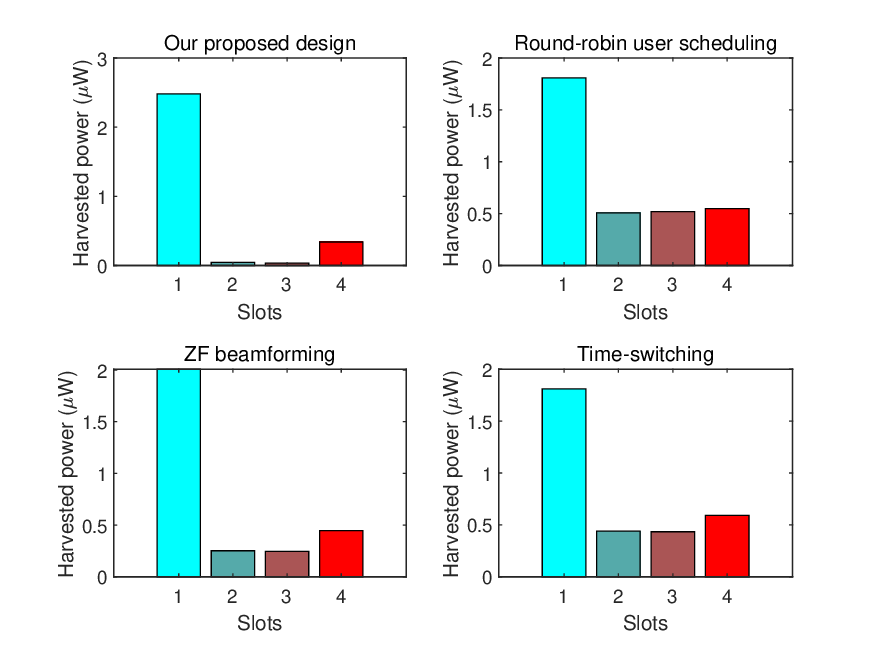}}
	\caption{Beampattern matching performance and the resultant communication rates at IRs and harvested power at the ER, where high communication rate constraint \(\Gamma_\text{IR} = 200\) Kbps and low harvested power constraint \(\Gamma_\text{ER} = 1\) \(\mu\)W are considered.}
	\label{Fig_HL}
\vspace{-0.1cm}\end{figure*}
\begin{figure*}[htb!]
	\centering 
	\subfigure[Beampattern matching at each slot.] {\includegraphics[width=0.64\columnwidth]{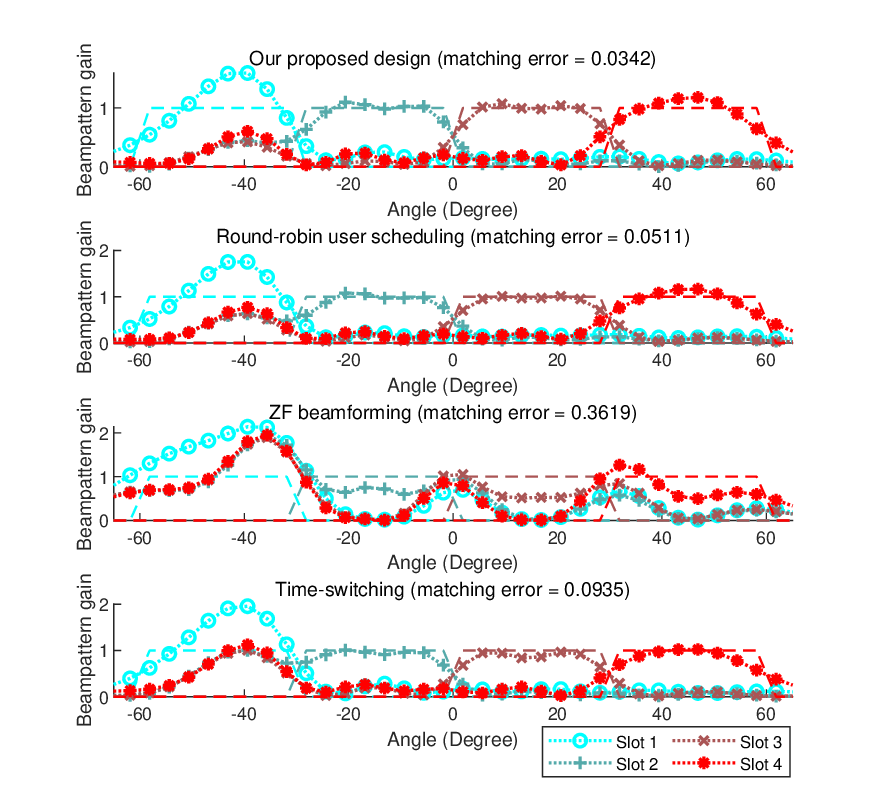}}
	\subfigure[Communication rate of each IR at each slot.] {\includegraphics[width=0.68\columnwidth]{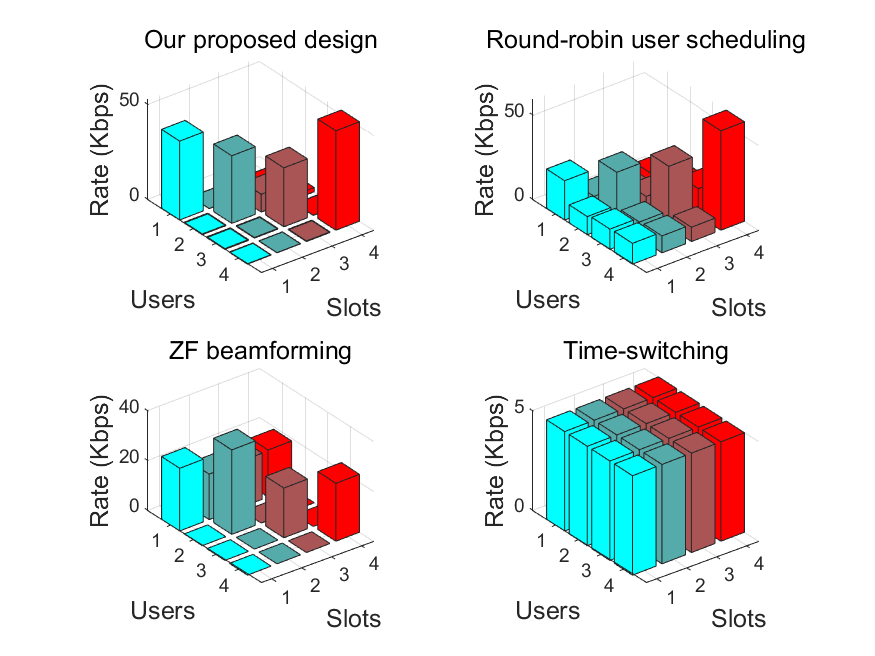}}
	\subfigure[Harvested power of the ER at each slot.] {\includegraphics[width=0.68\columnwidth]{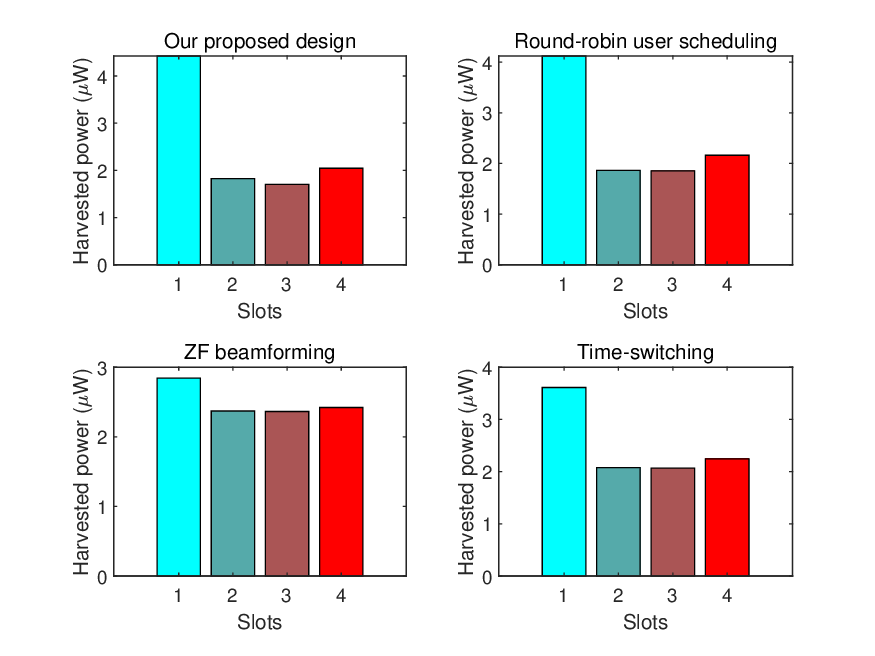}}
	\caption{Beampattern matching performance and the resultant communication rates at IRs and harvested power at the ER, where low communication rate constraint \(\Gamma_\text{IR} = 20\) Kbps and high harvested power constraint \(\Gamma_\text{ER} = 10\) \(\mu\)W are considered.}
	\label{Fig_LH}
\vspace{-0.1cm}\end{figure*}

Fig. \ref{Fig_LL} shows the beampattern matching performance and the resultant communication rates at IRs and harvested power at the ER under low communication rate constraint \(\Gamma_\text{IR} = 20\) Kbps and low harvested power constraint \(\Gamma_\text{ER} = 1\) \(\mu\)W.
As shown in Fig. \ref{Fig_LL}(a), it is observed that our proposed design as well as round-robin user scheduling and time-switching designs achieve excellent matching to the desired beampattern, while the ZF beamforming performs slightly worse. This is due to the fact that the ZF beamforming is a communication-oriented design and has natural disadvantages in beampattern matching.
In addition, it is observed in Figs. \ref{Fig_LL}(b) and \ref{Fig_LL}(c) that both the communication rates and harvested power achieved by the four designs exceed the corresponding constraints \(\Gamma_\text{IR}\) and \(\Gamma_\text{ER}\). This indicates that the four designs try their best for beampattern matching, while the by-product can satisfy the communication and WPT requirements.
In this case, each IR/ER hardly receives information/energy from the BS unless it is steered by the scanning beam in a certain time slot.
%thereby minimizing interference with transmit beampatterns.

Fig. \ref{Fig_HL} shows the beampattern matching performance and the resultant communication rates at IRs and harvested power at the ER under high communication rate constraint \(\Gamma_\text{IR} = 200\) Kbps and low harvested power constraint \(\Gamma_\text{ER} = 1\) \(\mu\)W.
Compared with the results in Fig. \ref{Fig_LL}(a), it is observed in Fig. \ref{Fig_HL}(a) that increasing the communication requirements has negligible impact on the beampattern matching performance in our proposed design, while noticeably affecting those with round-robin user scheduling and time-switching.
This corresponds to the results in Fig. \ref{Fig_HL}(b), where it is observed that our proposed design and the communication-oriented ZF beamforming maintain similar scheduling of IRs as seen in Fig. \ref{Fig_LL}(b), thereby minimizing the distortion on beampattern matching. 
In contrast, the round-robin user scheduling and time-switching designs evenly allocate the communication resources across IRs over time slots, leading to beampattern distortions near IRs. This is because the round-robin user scheduling only invokes one IR per subcarrier, restricting its frequency band utilization efficiency due to the communication rate limit at each IR. 
%Indeed, if we further increase \(\Gamma_\text{IR}\), it will quickly become infeasible when each IR reaches its rate limit.
On the other hand, the time-switching design guarantees the minimum communication rate for all IRs, preventing from increasing the communication rate for a specific IR alone.
Moreover, as shown in Fig. \ref{Fig_HL}(c), for the four designs, the ER is mostly served in time slot \(1\), which is consistent with the beam scanning scheme.

Fig. \ref{Fig_LH} shows the beampattern matching performance and the resultant communication rates at IRs and harvested power at the ER under low communication rate constraint \(\Gamma_\text{IR} = 20\) Kbps and high harvested power constraint \(\Gamma_\text{ER} = 10\) \(\mu\)W.
Compared with the results in Fig. \ref{Fig_LL}(a), it is observed in Fig. \ref{Fig_LH}(a) that the four designs suffer beampattern distortion caused by the increased WPT requirements to different extent.
This corresponds to the results in Fig. \ref{Fig_LH}(c), where it is observed that compared to Fig. \ref{Fig_LL}(c), the ER needs to be powered in the four slots, since powering the ER merely in slot \(1\) (consistent with the sensing beam scanning at this slot) is insufficient to meet the increased WPT requirements, thus leading to beampattern distortions near the ER.
For instance, since the ZF beamforming is oriented to communication, it struggles to make the transmitted power be harvested by the ER in slot \(1\). As a result, more power needs to be transmitted to the ER in slots \(2\sim4\), which severely exacerbates beampattern mismatch and leads to the high communication rate of IR \(1\) in Fig. \ref{Fig_LH}(b).
In contrast, leveraging high DoFs in beamforming and resource allocation, our proposed design is shown to transmit as much power to the ER as possible in slot \(1\), thus outperforming the three benchmarks.

\begin{figure}[htb]
	\centering {\includegraphics[width=0.4\textwidth]{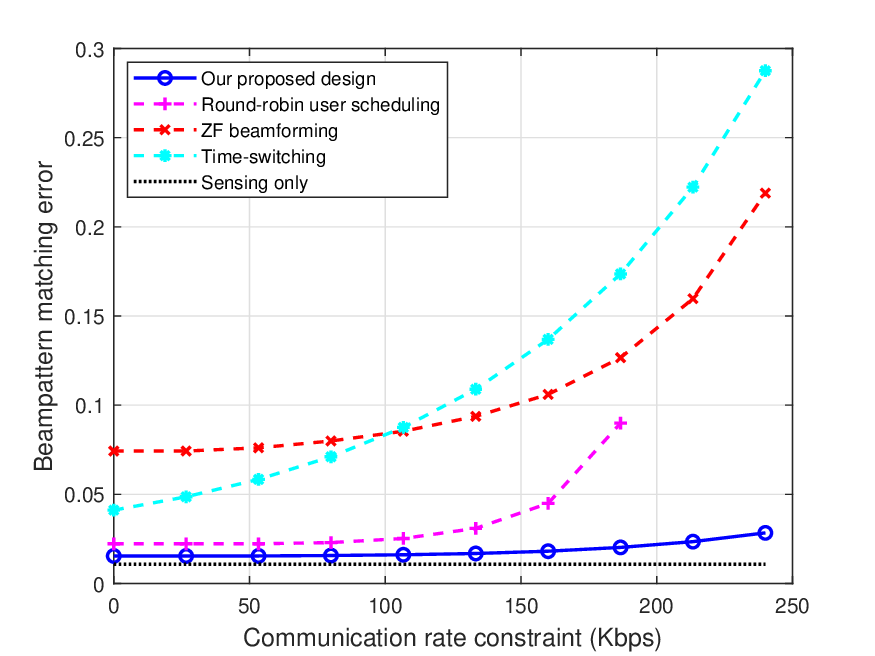}} 
	\caption{Beampattern matching error versus communication rate constraint \(\Gamma_\text{IR}\) with \(\Gamma_\text{ER} = 8\) \(\mu\)W.}
	\label{Fig_Gamc}
\vspace{-0.1cm}\end{figure}
\begin{figure}[htb]
	\centering {\includegraphics[width=0.4\textwidth]{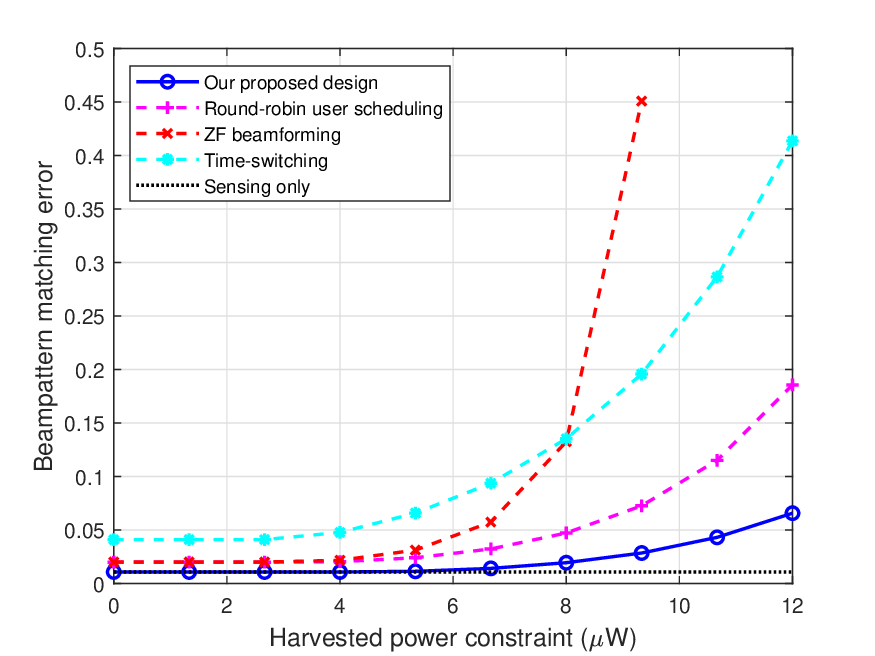}} 
	\caption{Beampattern matching error versus harvested power constraint \(\Gamma_\text{ER}\) with \(\Gamma_\text{IR} = 160\) Kbps.}
	\label{Fig_Gamp}
\vspace{-0.1cm}\end{figure}
\begin{figure}[htb!]
	\centering {\includegraphics[width=0.4\textwidth]{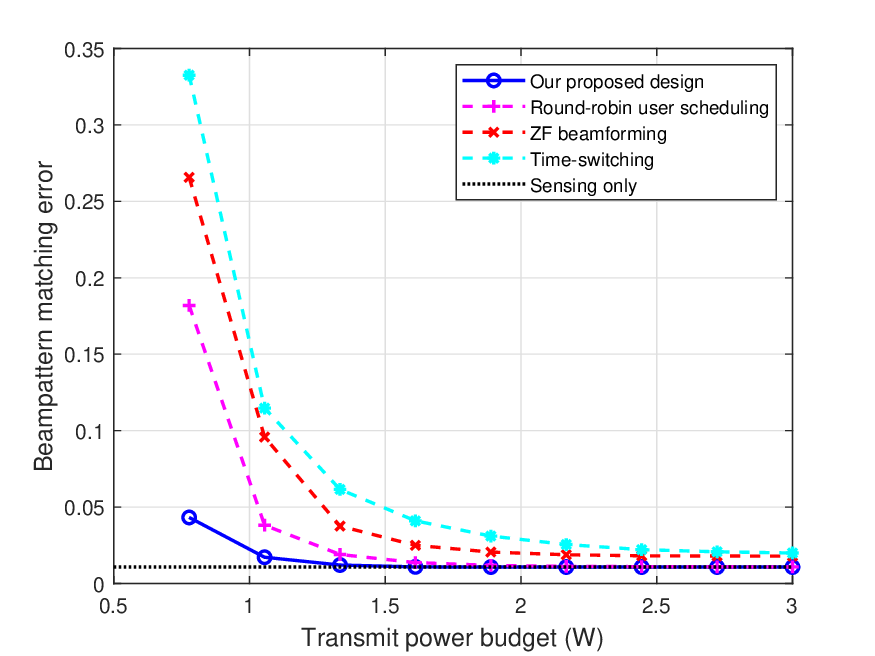}} 
	\caption{Beampattern matching error versus transmit power budget \(P_0\) with \(\Gamma_\text{IR} = 160\) Kbps and \(\Gamma_\text{ER} = 8\) \(\mu\)W.}
	\label{Fig_P0}
\vspace{-0.1cm}\end{figure}

Figs. \ref{Fig_Gamc}, \ref{Fig_Gamp}, and \ref{Fig_P0} show the achieved beampattern matching error versus the communication rate constraint \(\Gamma_\text{IR}\), harvested power constraint \(\Gamma_\text{ER}\), and transmit power budget \(P_0\), respectively.
It is observed that our proposed design performs close to the error lower bound achieved by sensing only and outperforms the three heuristic benchmarks under different constraints.
As shown in Fig. \ref{Fig_Gamc}, it is also observed that as \(\Gamma_\text{IR}\) increases, the round-robin user scheduling initially performs close to our proposed design and sensing only, but quickly becomes infeasible. This can be explained similarly as for Fig. \ref{Fig_HL}. In contrast, although the ZF beamforming and time-switching designs exhibit poorer performance at low \(\Gamma_\text{IR}\), they remain feasible at high \(\Gamma_\text{IR}\).
In addition, it is observed in Fig. \ref{Fig_Gamp} that as \(\Gamma_\text{ER}\) increases, the performance of ZF beamforming deteriorates more rapidly than the other designs, for similar reasons explained in Fig. \ref{Fig_LH}.
Moreover, it is observed in Fig. \ref{Fig_P0} that as \(P_0\) increases, the performance of round-robin user scheduling quickly approaches the error lower bound achieved by sensing only, while there remains a performance gap for ZF beamforming and time-switching at high \(P_0\).

\begin{figure}[tb]
	\centering {\includegraphics[width=0.4\textwidth]{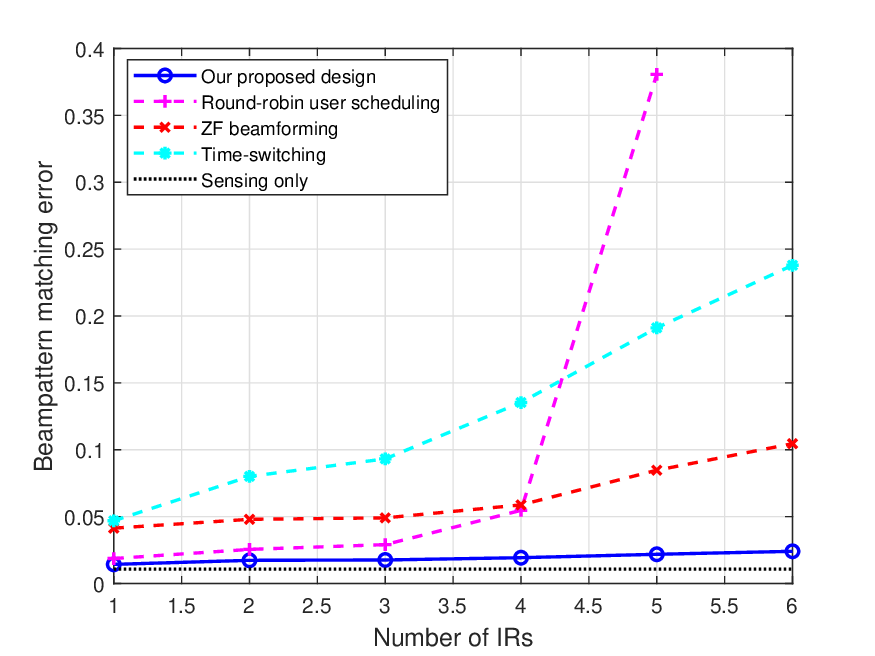}} 
	\caption{Beampattern matching error versus the number of IRs \(K_\text{IR}\) with \(\Gamma_\text{IR} = 160\) Kbps and \(\Gamma_\text{ER} = 8\) \(\mu\)W.}
	\label{Fig_K}
\vspace{-0.1cm}\end{figure}
\begin{figure}[tb!]
	\centering {\includegraphics[width=0.4\textwidth]{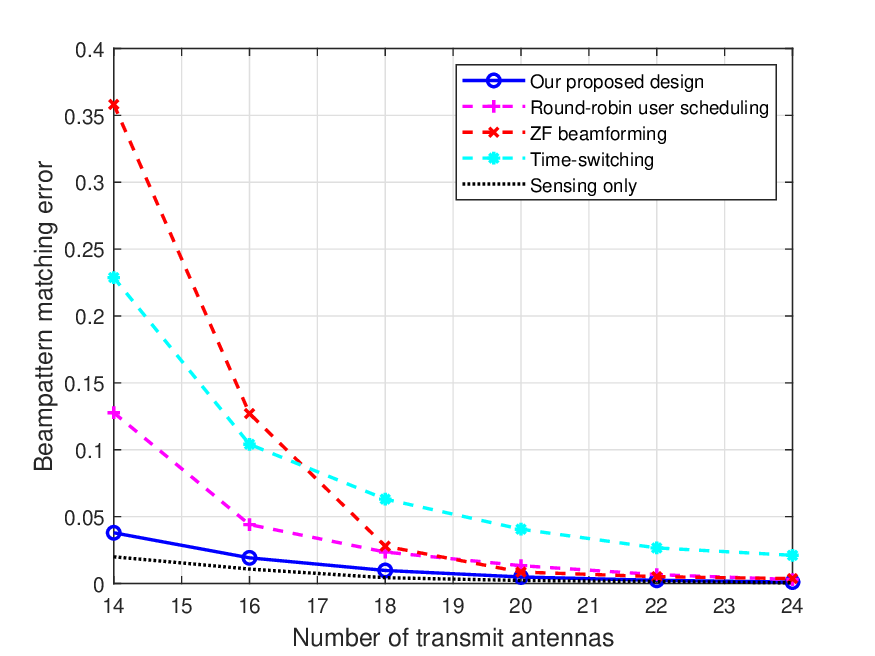}} 
	\caption{Beampattern matching error versus the number of transmit antennas \(N_t\) with \(\Gamma_\text{IR} = 160\) Kbps and \(\Gamma_\text{ER} = 8\) \(\mu\)W.}
	\label{Fig_N}
\vspace{-0.1cm}\end{figure}

Figs. \ref{Fig_K} and \ref{Fig_N} show the achieved beampattern matching error versus the number of IRs \(K_\text{IR}\) and the number of transmit antennas \(N_t\), respectively, where \(\Gamma_\text{IR} = 160\) Kbps and \(\Gamma_\text{ER} = 8\) \(\mu\)W. It is observed that our proposed design performs close to the lower bound achieved by sensing only and outperforms the three heuristic benchmarks under different system setups. It is also observed that the beampattern matching error achieved by each design increases as \(K_\text{IR}\) increases, and decreases as \(N_t\) increases. As shown in Fig. \ref{Fig_K}, when \(K_\text{IR}\) increases, the communication requirements become high, leading to rapid sensing performance deterioration for round-robin user scheduling. This is consistent with the results in Fig. \ref{Fig_Gamc}. In addition, as shown in Fig. \ref{Fig_N}, small \(N_t\) results in low transmit power gain towards the ER and equivalently increases the WPT requirements, making the ZF beamforming perform worse than the other designs. This is consistent with the results in Fig. \ref{Fig_Gamp}.

\begin{figure}[tb]
	\centering 
	\subfigure[Case with high communication rate constraint \(\Gamma_\text{IR} = 200\) Kbps and low harvested power constraint \(\Gamma_\text{ER} = 5\) \(\mu\)W.]
	{\includegraphics[width=0.8\columnwidth]{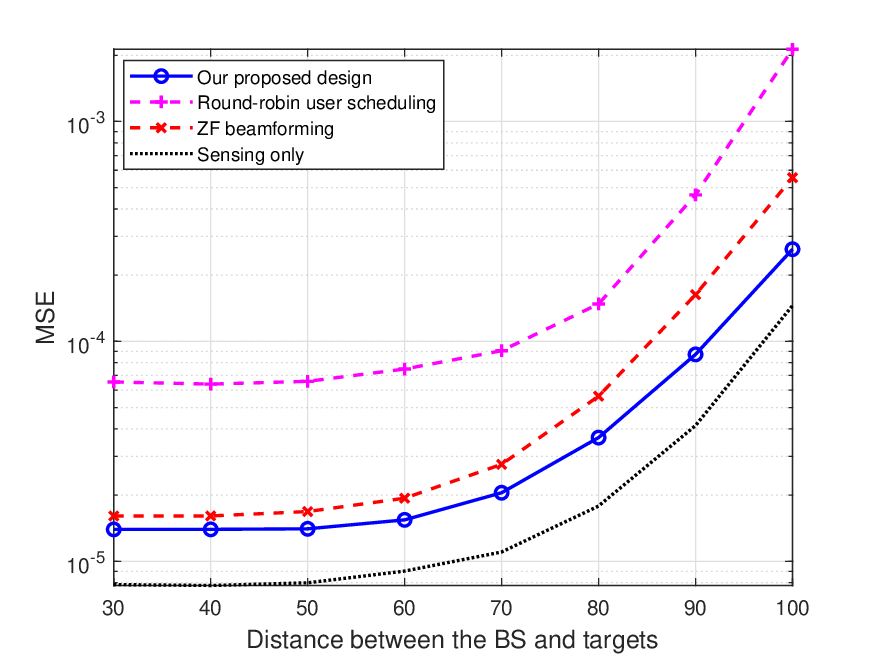}}
	\subfigure[Case with low communication rate constraint \(\Gamma_\text{IR} = 100\) Kbps and high harvested power constraint \(\Gamma_\text{ER} = 10\) \(\mu\)W.]
	{\includegraphics[width=0.8\columnwidth]{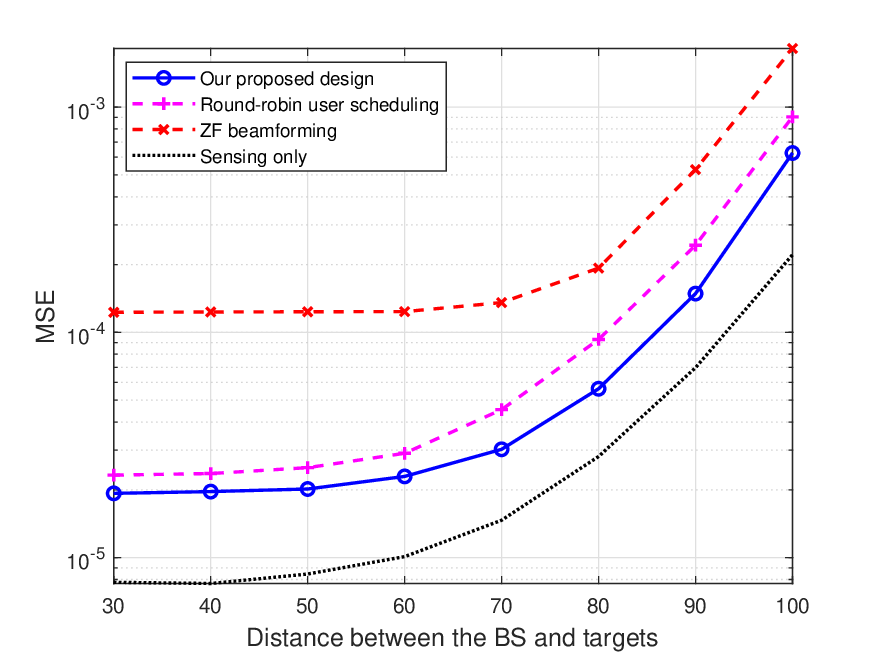}}
	\caption{MSE for target angle estimation versus the average distance between the BS and targets.}
	\label{Fig_mse}
\vspace{-0.1cm}\end{figure}
Finally, we show the real estimation performance of the proposed transmit beamforming designs, by considering specific target estimation tasks, to validate the effectiveness of our beampattern matching for sensing. The adopted angle, delay, and Doppler shift estimation methods are introduced in Appendix B.
Here, we only focus on the target angle estimation performance, as the estimated angles can be utilized for subsequent signal processing and will directly affect the accuracy of delay and Doppler shift estimation \cite{islam2022integrated, liu2020joint, pucci2022system, wei2023integrated}.
In the simulation, there is a set \(\mathcal{K}_\text{S} = \{1,\dots,8\}\) of point targets uniformly located in the range of \([-60^\circ, 60^\circ]\). The mean square error (MSE) is calculated by \(\mathrm{MSE} = \frac{1}{|\mathcal{K}_\text{S}|} \sum_{k\in\mathcal{K}_\text{S}} |\theta_k - \hat{\theta}_k|^2\), where \(\theta_k\) and \(\hat{\theta}_k\) denote the ground-truth and estimated angles of each target \(k \in\mathcal{K}_\text{S}\), respectively. 
Fig. \ref{Fig_mse} shows the MSE for target angle estimation versus the average distance between the BS and targets. It is observed that with the same estimation signal processing at the receiver, our proposed design achieves more accurate estimation results than the ZF beamforming and round-robin user scheduling designs.
As shown in Fig. \ref{Fig_mse}(a), under high \(\Gamma_\text{IR}\), the round-robin user scheduling performs worse than the ZF beamforming. This is consistent with the result in Fig. \ref{Fig_HL}(a), where the beampattern of round-robin user scheduling suffers from high distortion near angles of IRs. During beam scanning, the received signal in each slot is severely interfered by the echoes from uninterested angles, leading to inaccurate estimation in the current scanning area. 
In contrast, as shown in Fig. \ref{Fig_mse}(b), under high \(\Gamma_\text{ER}\), the estimation performance of ZF beamforming is worse than that of round-robin user scheduling, which can be explained by the large beampattern distortion near the angle of the ER shown in Fig. \ref{Fig_LH}(a).

\section{Conclusion}

This paper studied the joint transmit beamforming design for a multi-antenna OFDM ISCAP system. We minimized the beampattern matching error of beam scanning for sensing, subject to communication rate requirements at IRs and harvested power requirements at ERs, respectively. We proposed optimization algorithms to solve the formulated ISCAP problem, via the techniques of SDR, SCA, and FP. Our proposed designs exhibit dynamic resource allocation and user scheduling among time-frequency-space domains, thereby balancing the performance tradeoff among the three functions, and outperforming the heuristic benchmarks including ZF beamforming, round-robin user scheduling, and time switching.
It is our hope that this paper can provide new insights on user scheduling and multi-dimensional resource allocation for emerging multi-functional wireless networks towards 6G. 

\appendix

\subsection{Proof of Proposition 1}

%For each \(n\in\mathcal{N}\) and \(l\in\mathcal{L}\), we construct the projected solution \(\big(\{\bar{\mathbf{W}}_{n,l,k}\}, \bar{\zeta}\big)\) as 

First, it follows from \eqref{trans2} that \(\mathrm{rank}(\bar{\mathbf{W}}_{n,l,k}) \le 1\) and \(\bar{\mathbf{W}}_{n,l,k} \succeq \mathbf{0}\), \(\forall n\in\mathcal{N}, l\in\mathcal{L}, k\in\mathcal{K}_\text{IR}\). According to the Cauchy-Schwarz inequality, for any vector \(\mathbf{p} \in \mathbb{C}^{N_t\times 1}\), it holds that \(\mathbf{p}^H (\hat{\mathbf{W}}_{n,l,k} - \bar{\mathbf{W}}_{n,l,k}) \mathbf{p} = \mathbf{p}^H \hat{\mathbf{W}}_{n,l,k} \mathbf{p} - \frac{|\mathbf{p}^H \hat{\mathbf{W}}_{n,l,k} \mathbf{h}_{n,k}|^2}{\mathbf{h}_{n,k}^H \hat{\mathbf{W}}_{n,l,k} \mathbf{h}_{n,k}} \ge 0\), i.e., \(\hat{\mathbf{W}}_{n,l,k} \succeq \bar{\mathbf{W}}_{n,l,k}\), \(\forall k\in\mathcal{K}_\text{IR}\). Consequently, we have \(\bar{\mathbf{W}}_{n,l,0} = \hat{\mathbf{W}}_{n,l,0} + \sum_{k\in\mathcal{K}_\text{IR}} (\hat{\mathbf{W}}_{n,l,k} - \bar{\mathbf{W}}_{n,l,k}) \succeq \mathbf{0}\), and therefore, \(\{\bar{\mathbf{W}}_{n,l,k}\}\) and \(\bar{\zeta}\) satisfy constraint \eqref{1sdp}.

Next, according to \eqref{trans3}, we have \(\sum_{k\in\{0\}\cup\mathcal{K}_\text{IR}} \bar{\mathbf{W}}_{n,l,k} = \sum_{k\in\{0\}\cup\mathcal{K}_\text{IR}} \hat{\mathbf{W}}_{n,l,k}\). Therefore, the objective function \eqref{1s} and the LHSs of constraints \eqref{1p} and \eqref{1b} achieved by \(\{\bar{\mathbf{W}}_{n,l,k}\}\) and \(\bar{\zeta}\) remain the same as that by \(\{\hat{\mathbf{W}}_{n,l,k}\}\) and \(\hat{\zeta}\). 

Then, by substituting \eqref{trans2} and \eqref{trans3} into constraint \eqref{1c}, we verify that \(\mathbf{h}_{n,k}^H \bar{\mathbf{W}}_{n,l,k} \mathbf{h}_{n,k} = \mathbf{h}_{n,k}^H \hat{\mathbf{W}}_{n,l,k} \mathbf{h}_{n,k}\) and \(\mathbf{h}_{n,k}^H (\sum_{i\in\{0\}\cup\mathcal{K}_\text{IR}\setminus\{k\}} \bar{\mathbf{W}}_{n,l,i}) \mathbf{h}_{n,k} = \mathbf{h}_{n,k}^H (\sum_{i\in\{0\}\cup\mathcal{K}_\text{IR}\setminus\{k\}} \hat{\mathbf{W}}_{n,l,i} + \hat{\mathbf{W}}_{n,l,k} - \bar{\mathbf{W}}_{n,l,k}) \mathbf{h}_{n,k} = \mathbf{h}_{n,k}^H (\sum_{i\in\{0\}\cup\mathcal{K}_\text{IR}\setminus\{k\}} \hat{\mathbf{W}}_{n,l,i}) \mathbf{h}_{n,k}\), \(\forall k\in\mathcal{K}_\text{IR}\). As a result, the LHS of constraint \eqref{1c} achieved by \(\{\bar{\mathbf{W}}_{n,l,k}\}\) and \(\bar{\zeta}\) remains the same as that by \(\{\hat{\mathbf{W}}_{n,l,k}\}\) and \(\hat{\zeta}\). Thus, this completes the proof.

\subsection{Estimation Signal Processing}

To evaluate the performance of the proposed transmit beamforming designs for practical target estimation tasks, in the following we present practical signal processing for estimating the target parameters including angle, delay, and Doppler shift.

In slot \(q\in\mathcal{Q}\), we assume that there is a set \(\mathcal{K}_{\text{S},q}\) of \(K_{\text{S},q}\) point targets within the interested angle range.
Let \(\bar{\mathbf{H}}_k = b_k \mathbf{a}(\theta_k) \mathbf{v}(\theta_k)^T\) denote the target response matrix from the BS transmitter to each target \(k\in\mathcal{K}_{\text{S},q}\) to the BS receiver, where \(b_k\) denotes the complex  amplitude, \(\theta_k\) denotes the direction of arrival (DoA), and \(\mathbf{a}(\theta_k)\) and \(\mathbf{v}(\theta_k)\) denote the receive and transmit array steering vectors w.r.t. \(\theta_k\), respectively. Furthermore, we denote \(\tau_k\) as the round-trip delay associated with target \(k\) and \(\nu_k = \frac{2v_k}{c}\) as the corresponding normalized Doppler shift, where \(v_k\) and \(c\) denote the radial velocity and the speed of light, respectively.

Based on the extended Saleh-Valenzuela channel model \cite{rappaport2015millimeter}, the received signal after removing the cyclic prefix (CP)  \cite{keskin2021limited} at each symbol \(l\in\mathcal{L}\) and subcarrier \(n\in\mathcal{N}\) is given by
\begin{equation}
	\mathbf{y}_{n,l} = \sum_{k\in\cup_{q\in\mathcal{Q}}\mathcal{K}_{\text{S},q}} e^{j 2\pi (l \nu_k f_c T_\text{sym} - n \tau_k \Delta_f)}\bar{\mathbf{H}}_k \mathbf{x}_{n,l} + \mathbf{z}_{n,l},
\end{equation}
where \(\mathbf{z}_{n,l} \in \mathbb{C}^{N_r\times 1}\) denotes the noise at the BS receiver that is a CSCG random vector with zero mean and covariance \(\sigma_s^2 \mathbf{I}_{N_r}\). Based on the received signals \(\{\mathbf{y}_{n,l}\}\), the BS aims to estimate the unknown parameters of each target \(k\), including DoA \(\theta_k\), delay \(\tau_k\), and Doppler shift \(\nu_k\).

\subsubsection{DoA Estimation}

During each time slot \(q\in\mathcal{Q}\), the sample covariance matrix of the received signals is given by
\begin{equation}
	\mathbf{R}_{y} = \sum_{l\in\mathcal{L}_q} \sum_{n\in\mathcal{N}} \mathbf{y}_{n,l} \mathbf{y}_{n,l}^H.
\end{equation}
For DoA estimation, we use the multiple signal classification (MUSIC) algorithm \cite{schmidt1986multiple}. The MUSIC algorithm starts by performing EVD on \(\mathbf{R}_{y}\), i.e.,
\begin{equation}
	\mathbf{R}_{y} = \begin{bmatrix}
		\mathbf{U}_{y,1} & \mathbf{U}_{y,0}
	\end{bmatrix} \mathbf{\Lambda}_y \begin{bmatrix}
		\mathbf{U}_{y,1}^H \\ \mathbf{U}_{y,0}^H
	\end{bmatrix},
\end{equation}
where \(\mathbf{\Lambda}_y = \mathrm{diag}(\lambda_{y,1}, \dots, \lambda_{y,N_r})\) is a diagonal matrix with eigenvalues sorted in a descending order, i.e., \(\lambda_{y,1} \ge \dots \ge \lambda_{y,N_r}\), and \(\mathbf{U}_{y,1} \in \mathbb{C}^{N_r \times K_{\text{S},q}}\) and \(\mathbf{U}_{y,0} \in \mathbb{C}^{N_r \times (N_r - K_{\text{S},q})}\) contain the eigenvectors corresponding to largest \(K_{\text{S},q}\) and smallest \(N_r - K_{\text{S},q}\) eigenvalues, respectively. 
Then, the MUSIC spectrum w.r.t. the interested angles is formulated as
\begin{equation}
	S(\theta) = \frac{1}{\mathbf{a}^H(\theta) \mathbf{U}_{y,0} \mathbf{U}_{y,0}^H \mathbf{a}(\theta)},
\end{equation}
whose \(K_{\text{S},q}\) peaks are obtained as the estimated DoAs, denoted by \(\{\hat{\theta}_k\}_{k\in\mathcal{K}_{\text{S},q}}\).

\subsubsection{Delay and Doppler Shift Estimation}

Given the estimated DoAs and transmitted signal at each OFDM symbol \(l\in\mathcal{L}\) and subcarrier \(n\in\mathcal{N}\), we define the reference signal for each target \(k\in\cup_{q\in\mathcal{Q}}\mathcal{K}_{\text{S},q}\) as
\begin{equation}
	g_{n,l,k} = \mathbf{v}^T(\hat{\theta}_k) \mathbf{x}_{n,l}.
\end{equation}
%Next, we utilize the received signal \(\mathbf{y}_{n,l}\) to derive the quotient averaged across all receive antennas that includes the effect of delay and Doppler shift in the direction of \(\hat{\theta}_k\).
%\begin{equation}
%	c_{n,l,k} = \frac{1}{N_r} \sum_{i=1}^{N_r} \frac{[\mathbf{y}_{n,l}]_i}{[\mathbf{g}_{n,l,k}]_i} / \textcolor{blue}{c_{n,l,k} = \frac{\mathbf{y}_{n,l}}{\mathbf{g}_{n,l,k}}} / \textcolor{blue}{c_{n,l} = \frac{\mathbf{y}_{n,l}}{\mathbf{x}_{n,l}}}.
%\end{equation}
Notice that when the number of receive antennas is large, we have \(\mathbf{a}^H(\theta_i) \mathbf{a}(\theta_i) = N_r\) and \(\mathbf{a}^H(\theta_i) \mathbf{a}(\theta_j) \approx 0\), \(\forall \theta_i \ne \theta_j\). As such, the received signal can be processed as \cite{islam2022integrated, liu2020joint, pucci2022system, wei2023integrated}
\begin{equation}
	c_{n,l,k} = \mathbf{a}^H(\hat{\theta}_k) \mathbf{y}_{n,l} / g_{n,l,k} \approx e^{j 2\pi (l \nu_k f_c T_\text{sym}- n \tau_k \Delta_f)} N_r b_k.
\end{equation}
For each target \(k\), we formulate the likelihood function as
\begin{equation}
	A_k(i,j) = \sum_{l\in\mathcal{L}} \sum_{n\in\mathcal{N}} c_{n,l,k} e^{j 2\pi \frac{n i}{N}} e^{-j 2\pi \frac{l j}{L}}, \forall i,j \in \mathcal{N},\mathcal{L}.
\end{equation}
Then, we find the best quantized delay and Doppler shift as the ones that maximize the likelihood function norm, i.e.,
\begin{equation}
	(\hat{i}_k,\hat{j}_k) = \arg \max_{i,j \in \mathcal{N},\mathcal{L}} \big\{|A_k(i,j)|^2\big\}.
\end{equation}
Accordingly, the delay and Doppler shift of target \(k\) are estimated as \(\hat{\tau}_k = \frac{\hat{i}_k}{N \Delta_f}\) and \(\hat{\nu}_k = \frac{\hat{j}_k}{L f_c T_\text{sym}}\), respectively.

\bibliographystyle{IEEEtran}
\bibliography{ofdm-ISCAP_arxiv}

\begin{thebibliography}{10}
\providecommand{\url}[1]{#1}
\csname url@samestyle\endcsname
\providecommand{\newblock}{\relax}
\providecommand{\bibinfo}[2]{#2}
\providecommand{\BIBentrySTDinterwordspacing}{\spaceskip=0pt\relax}
\providecommand{\BIBentryALTinterwordstretchfactor}{4}
\providecommand{\BIBentryALTinterwordspacing}{\spaceskip=\fontdimen2\font plus
\BIBentryALTinterwordstretchfactor\fontdimen3\font minus
  \fontdimen4\font\relax}
\providecommand{\BIBforeignlanguage}[2]{{%
\expandafter\ifx\csname l@#1\endcsname\relax
\typeout{** WARNING: IEEEtran.bst: No hyphenation pattern has been}%
\typeout{** loaded for the language `#1'. Using the pattern for}%
\typeout{** the default language instead.}%
\else
\language=\csname l@#1\endcsname
\fi
#2}}
\providecommand{\BIBdecl}{\relax}
\BIBdecl

\bibitem{tong20226g}
W.~Tong and P.~Zhu, \emph{{6G: The Next Horizon}}.\hskip 1em plus 0.5em minus
  0.4em\relax Cambridge, U.K.: Cambridge Univ. Press, 2022.

\bibitem{cui2021integrating}
Y.~Cui, F.~Liu, X.~Jing, and J.~Mu, ``Integrating sensing and communications
  for ubiquitous {IoT}: Applications, trends, and challenges,'' \emph{IEEE
  Netw.}, vol.~35, no.~5, pp. 158--167, Sep. 2021.

\bibitem{liu2022Integrated}
F.~Liu, Y.~Cui, C.~Masouros, J.~Xu, T.~X. Han, Y.~C. Eldar, and S.~Buzzi,
  ``Integrated sensing and communications: Toward dual-functional wireless
  networks for {6G} and beyond,'' \emph{IEEE J. Sel. Areas Commun.}, vol.~40,
  no.~6, pp. 1728--1767, Jun. 2022.

\bibitem{zeng2017communications}
Y.~Zeng, B.~Clerckx, and R.~Zhang, ``Communications and signals design for
  wireless power transmission,'' \emph{IEEE Trans. Commun.}, vol.~65, no.~5,
  pp. 2264--2290, May 2017.

\bibitem{clerckx2019fundamentals}
B.~Clerckx, R.~Zhang, R.~Schober, D.~W.~K. Ng, D.~I. Kim, and H.~V. Poor,
  ``Fundamentals of wireless information and power transfer: From {RF} energy
  harvester models to signal and system designs,'' \emph{IEEE J. Sel. Areas
  Commun.}, vol.~37, no.~1, pp. 4--33, Jan. 2019.

\bibitem{chen2024integrated}
Y.~Chen, Z.~Ren, J.~Xu, Y.~Zeng, D.~W.~K. Ng, and S.~Cui, ``Integrated sensing,
  communication, and powering ({ISCAP}): Towards multi-functional {6G} wireless
  networks,'' \emph{arXiv preprint arXiv:2401.03516}, 2024.

\bibitem{hua2024mimo}
H.~Hua, T.~X. Han, and J.~Xu, ``{MIMO} integrated sensing and communication:
  {CRB}-rate tradeoff,'' \emph{IEEE Trans. Wireless Commun.}, vol.~23, no.~4,
  pp. 2839--2854, Apr. 2024.

\bibitem{ren2024fundamental}
Z.~Ren, Y.~Peng, X.~Song, Y.~Fang, L.~Qiu, L.~Liu, D.~W.~K. Ng, and J.~Xu,
  ``Fundamental {CRB}-rate tradeoff in multi-antenna {ISAC} systems with
  information multicasting and multi-target sensing,'' \emph{IEEE Trans.
  Wireless Commun.}, vol.~23, no.~4, pp. 3870--3885, Apr. 2024.

\bibitem{xu2014multiuser}
J.~Xu, L.~Liu, and R.~Zhang, ``Multiuser {MISO} beamforming for simultaneous
  wireless information and power transfer,'' \emph{IEEE Trans. Signal
  Process.}, vol.~62, no.~18, pp. 4798--4810, Sep. 2014.

\bibitem{chen2024isac}
Y.~Chen, H.~Hua, J.~Xu, and D.~W.~K. Ng, ``{ISAC} meets {SWIPT}:
  Multi-functional wireless systems integrating sensing, communication, and
  powering,'' \emph{IEEE Trans. Wireless Commun.}, vol.~23, no.~8, pp.
  8264--8280, Aug. 2024.

\bibitem{zhou2023integrating}
Z.~Zhou, X.~Li, G.~Zhu, J.~Xu, K.~Huang, and S.~Cui, ``Integrating sensing,
  communication, and power transfer: Multiuser beamforming design,'' \emph{IEEE
  J. Sel. Areas Commun., Early Access}, Jun. 2024.

\bibitem{hao2024energy}
Z.~Hao, Y.~Fang, X.~Yu, J.~Xu, L.~Qiu, L.~Xu, and S.~Cui, ``Energy-efficient
  hybrid beamforming with dynamic on-off control for integrated sensing,
  communications, and powering,'' \emph{arXiv preprint arXiv:2403.16353}, 2024.

\bibitem{zhang2024integrated}
X.~Zhang and Q.~Zhu, ``Integrated sensing, communications, and powering for
  statistical-{QoS} provisioning over next-generation massive-{MIMO} mobile
  networks,'' in \emph{Proc. 2024 58th Annual Conference on Information
  Sciences and Systems (CISS)}, Mar. 2024, pp. 1--6.

\bibitem{ren2024secure}
Z.~Ren, S.~Zhang, X.~Li, L.~Qiu, J.~Xu, and D.~W.~K. Ng, ``Secure
  communications in near-filed {ISCAP} systems with extremely large-scale
  antenna arrays,'' in \emph{Proc. IEEE ISWCS 2024}, Jul. 2024, pp. 1--6.

\bibitem{yang2024joint}
Y.~Yang, H.~Gao, X.~Yang, R.~Cao, and Y.~Fan, ``Joint beamforming for
  {RIS}-assisted integrated communication, sensing and power transfer
  systems,'' \emph{IEEE Wireless Commun. Lett.}, vol.~13, no.~2, pp. 288--292,
  Feb. 2024.

\bibitem{li2023intelligent}
Z.~Li, Z.~Zhu, Z.~Chu, Y.~Guan, D.~Mi, F.~Liu, and L.-L. Yang, ``Intelligent
  reflective surface assisted integrated sensing and wireless power transfer,''
  \emph{IEEE Trans. Intell. Transp. Syst., Early Access}, May 2024.

\bibitem{mayer2023joint}
K.~M. Mayer, N.~Shanin, Z.~You, S.~Lotter, S.~Br{\"u}ckner, M.~Vossiek,
  L.~Cottatellucci, and R.~Schober, ``Joint transmit signal and beamforming
  design for integrated sensing and power transfer systems,'' in \emph{Proc.
  IEEE ICC 2024}, Jun. 2024, pp. 5086--5091.

\bibitem{li2022wirelessly}
X.~Li, Z.~Han, Z.~Zhou, Q.~Zhang, K.~Huang, and Y.~Gong, ``Wirelessly powered
  integrated sensing and communication,'' in \emph{Proc. the 1st ACM MobiCom
  Workshop on Integrated Sensing and Communications Systems}, Oct. 2022, pp.
  1--6.

\bibitem{zhang2023training}
L.~Zhang, Y.~Fang, Z.~Ren, L.~Qiu, and J.~Xu, ``Training-free energy
  beamforming assisted by wireless sensing,'' in \emph{Proc. IEEE WCNC 2024},
  Apr. 2024, pp. 1--6.

\bibitem{xu2024sensing}
Y.~Xu, D.~Xu, and S.~Song, ``Sensing-assisted robust {SWIPT} for mobile energy
  harvesting receivers,'' \emph{arXiv preprint arXiv:2402.09976}, 2024.

\bibitem{zhang2019multibeam}
J.~A. Zhang, X.~Huang, Y.~J. Guo, J.~Yuan, and R.~W. Heath, ``Multibeam for
  joint communication and radar sensing using steerable analog antenna
  arrays,'' \emph{IEEE Trans. Veh. Technol.}, vol.~68, no.~1, pp. 671--685,
  Jan. 2019.

\bibitem{pucci2022system}
L.~Pucci, E.~Paolini, and A.~Giorgetti, ``System-level analysis of joint
  sensing and communication based on {5G} new radio,'' \emph{IEEE J. Sel. Areas
  Commun.}, vol.~40, no.~7, pp. 2043--2055, Jul. 2022.

\bibitem{fuhrmann2008transmit}
D.~R. Fuhrmann and G.~San~Antonio, ``Transmit beamforming for {MIMO} radar
  systems using signal cross-correlation,'' \emph{IEEE Trans. Aerosp. Electron.
  Syst.}, vol.~44, no.~1, pp. 171--186, Jan. 2008.

\bibitem{liu2018mumimo}
F.~Liu, C.~Masouros, A.~Li, H.~Sun, and L.~Hanzo, ``{MU-MIMO} communications
  with {MIMO} radar: From co-existence to joint transmission,'' \emph{IEEE
  Trans. Wireless Commun.}, vol.~17, no.~4, pp. 2755--2770, Apr. 2018.

\bibitem{zhang2023multi}
Y.~Zhang, S.~Aditya, and B.~Clerckx, ``Multi-functional {OFDM} signal design
  for integrated sensing, communications, and power transfer,'' \emph{arXiv
  preprint arXiv:2311.00104}, 2023.

\bibitem{liuxiang2020joint}
X.~Liu, T.~Huang, N.~Shlezinger, Y.~Liu, J.~Zhou, and Y.~C. Eldar, ``Joint
  transmit beamforming for multiuser {MIMO} communications and {MIMO} radar,''
  \emph{IEEE Trans. Signal Process.}, vol.~68, pp. 3929--3944, Jun. 2020.

\bibitem{telatar1999capacity}
E.~Telatar, ``Capacity of multi-antenna {Gaussian} channels,'' \emph{Eur.
  trans. telecommun.}, vol.~10, no.~6, pp. 585--595, Nov. 1999.

\bibitem{luo2010semidefinite}
Z.-Q. Luo, W.-K. Ma, A.~M.-C. So, Y.~Ye, and S.~Zhang, ``Semidefinite
  relaxation of quadratic optimization problems,'' \emph{IEEE Signal Process.
  Mag.}, vol.~27, no.~3, pp. 20--34, May 2010.

\bibitem{sun2017majorization}
Y.~Sun, P.~Babu, and D.~P. Palomar, ``Majorization-minimization algorithms in
  signal processing, communications, and machine learning,'' \emph{IEEE Trans.
  Signal Process.}, vol.~65, no.~3, pp. 794--816, Feb. 2017.

\bibitem{cvx2012}
I.~CVX~Research, ``{CVX}: Matlab software for disciplined convex programming,
  version 2.0,'' \url{https://cvxr.com/cvx}, Aug. 2012.

\bibitem{shen2018fractional}
K.~Shen and W.~Yu, ``Fractional programming for communication systems—{Part
  I}: Power control and beamforming,'' \emph{IEEE Trans. Signal Process.},
  vol.~66, no.~10, pp. 2616--2630, Mar. 2018.

\bibitem{shen2018fractional2}
------, ``Fractional programming for communication systems—{Part II}: Uplink
  scheduling via matching,'' \emph{IEEE Trans. Signal Process.}, vol.~66,
  no.~10, pp. 2631--2644, Mar. 2018.

\bibitem{islam2022integrated}
M.~A. Islam, G.~C. Alexandropoulos, and B.~Smida, ``Integrated sensing and
  communication with millimeter wave full duplex hybrid beamforming,'' in
  \emph{Proc. IEEE ICC 2022}, May 2022, pp. 4673--4678.

\bibitem{liu2020joint}
F.~Liu, C.~Masouros, A.~P. Petropulu, H.~Griffiths, and L.~Hanzo, ``Joint radar
  and communication design: Applications, state-of-the-art, and the road
  ahead,'' \emph{IEEE Trans. Commun.}, vol.~68, no.~6, pp. 3834--3862, Jun.
  2020.

\bibitem{wei2023integrated}
Z.~Wei, H.~Qu, Y.~Wang, X.~Yuan, H.~Wu, Y.~Du, K.~Han, N.~Zhang, and Z.~Feng,
  ``Integrated sensing and communication signals toward {5G-A} and {6G}: A
  survey,'' \emph{IEEE Int. Things J.}, vol.~10, no.~13, pp. 11\,068--11\,092,
  Jul. 2023.

\bibitem{rappaport2015millimeter}
T.~S. Rappaport, R.~W. Heath~Jr, R.~C. Daniels, and J.~N. Murdock,
  \emph{Millimeter wave wireless communications}.\hskip 1em plus 0.5em minus
  0.4em\relax Pearson Education, 2015.

\bibitem{keskin2021limited}
M.~F. Keskin, V.~Koivunen, and H.~Wymeersch, ``Limited feedforward waveform
  design for {OFDM} dual-functional radar-communications,'' \emph{IEEE Trans.
  Signal Process.}, vol.~69, pp. 2955--2970, Apr. 2021.

\bibitem{schmidt1986multiple}
R.~Schmidt, ``Multiple emitter location and signal parameter estimation,''
  \emph{IEEE Trans. Antennas Propagat.}, vol.~34, no.~3, pp. 276--280, Mar.
  1986.

\end{thebibliography}

\end{document}